\documentclass[preprint]{aastex}

\newcommand{\HST}{{\sl HST}}
\newcommand{\IUE}{{\sl IUE}}
\newcommand{\Msun}{$M_{\sun}$}
\newcommand{\degree}{\mbox{$^{\circ}$}}
\newcommand{\perone}{\mbox{$^{-1}$}}
\newcommand{\pertwo}{\mbox{$^{-2}$}}

\newcommand{\etal}{et al.}
\newcommand{\eg}{e.g.}
\newcommand{\ie}{i.e.}

\newcommand{\IRAS}{{\sl IRAS}}

\newcommand{\hii}{\ion{H}{2}}
\newcommand{\oii}{[\ion{O}{2}]~$\lambda$3727}
\newcommand{\mgii}{{\ion{Mg}{2}}~$\lambda$2800}
\newcommand{\oiin}{[\ion{O}{2}]}
\newcommand{\mgiin}{{\ion{Mg}{2}}}
\newcommand{\hf}{\mbox{$h_{50}$}}
\newcommand{\hfperone}{\mbox{$h_{50}^{-1}$}}
\newcommand{\kms}{\hbox{km~s$^{-1}$}}
\newcommand{\Rs}{\mbox{${\cal R}$}}   

\slugcomment{to appear in the June 2000 issue of the Astronomical Journal}

\shorttitle{An ERO Concentration at $z=1.31$}
\shortauthors{Liu et al.}

\begin{document}

\title{\large Extremely Red Objects in the Field of QSO~1213--0017:\\
A Galaxy Concentration at $z$ = 1.31\altaffilmark{1,2}}

\author{\sc Michael C. Liu} 
\affil{Department of Astronomy, University of California, Berkeley, CA 94720}
\email{mliu@astro.berkeley.edu}  

\author{\sc Arjun Dey\altaffilmark{3}}
\affil{National Optical Astronomy Observatories, Tucson, AZ 85719}

\author{\sc James R. Graham and Kevin A. Bundy}
\affil{Department of Astronomy, University of California,
    Berkeley, CA 94720}

\author{\sc Charles C. Steidel\altaffilmark{4} and Kurt Adelberger}
\affil{Palomar Observatory, Caltech 105-24, Pasadena, CA 91125}

\author{\sc Mark E. Dickinson}
\affil{Space Telescope Science Institute, 3700 San Martin Drive,
Baltimore, MD 21218}

\altaffiltext{1}{Based in part on observations obtained with the
NASA/ESA {\sl Hubble Space Telescope}, which is operated by the STScI
for the Association of Universities in Research in Astronomy, Inc.,
under NASA contract NAS5-26555.}

\altaffiltext{2}{Based in part on observations obtained at the
W. M. Keck Observatory, which is operated jointly by the California
Institute of Technology and the University of California.}

\altaffiltext{3}{Hubble Fellow}

\altaffiltext{4}{Visiting Astronomer, Kitt Peak National Observatories
(KPNO), a division of National Optical Astronomy Observatories, which is
operated by Associated Universities for Research in Astronomy, Inc.,
under cooperative agreement with the National Science Foundation.}

\begin{abstract}
\noindent We have discovered a concentration of extremely red objects
(EROs; $R-K>6$) in the field of the \mbox{$z=2.69$} quasar
QSO~1213--0017 (UM~485), which is significantly overabundant compared to
the field ERO surface density.  The optical/near-IR colors of the EROs
and numerous other red galaxies in this field are consistent with
elliptical galaxies at $z=1-2$.  \HST\ optical images for a subset of
galaxies show regular morphologies, most of them being disky or diffuse
and without any obvious evidence for interactions. Ground-based IR
images show similar morphologies, indicating any dust reddening in these
objects is spatially uniform. Optical spectroscopy with the W. M. Keck
Telescope has found that four of the red galaxies lie at $z\approx1.31$,
and a fifth lies in the foreground at $z=1.20$. Of the $z\approx1.31$
galaxies, one is a reddened AGN while the remaining three have
rest-frame UV absorption-line spectra characteristic of old (few Gyr)
stellar populations, similar to the old red galaxy LBDS~53W091 at
$z=1.55$.  Including the \mgiin\ absorber seen in the QSO spectrum, we
find five galaxies at $z\approx1.31$ spread over $1.5~\hfperone$~Mpc on
the sky.  These results suggest we have discovered a coherent structure
of old galaxies at high-redshift, possibly associated with a massive
galaxy cluster.
\end{abstract}


\section{Introduction}

The observational study of galaxy formation and evolution can broadly be
divided into two approaches: (1) searches for primordial galaxies,
typically oriented toward high redshifts, in order to scrutinize
formation and evolution processes as they occur, and (2) studies of
existing old galaxies to decipher their origin and life history from
their current properties.  With the advent of infrared imaging
detectors, a population of infrared-bright, extremely red objects (EROs)
have been uncovered which is relevant to both of these approaches.

\citet{1988ApJ...331L..77E} discovered a few objects with exceptional
optical/IR colors ($R-K\gtrsim5$) in the first deep near-IR sky survey.
While the brightest objects were shown to be $z\leq0.8$ ellipticals
\citep{1988ApJ...332L..59L,1989ApJ...341...80E}, their tantalizing
conjecture that the fainter ($K\gtrsim18$) red objects are $z>1$
ellipticals has remained unresolved until recently.  Subsequent deep
infrared imaging of primarily high-redshift radio galaxy and quasar
fields serendipitously uncovered handfuls of objects with extreme colors
\citep{1992ApJ...386...52M,1992ApJ...399L..47E,1994ApJ...420L...5G,
1994AJ....107.1303H,1994ApJ...420L...1S,1995ApJ...440..515D,
1995ApJ...438L..13D}.  Deep multicolor optical/IR sky surveys have begun
to assemble larger samples of these objects using well-defined selection
criteria, an essential starting point for statistical studies
\citep{1994ApJ...434..114C,1998ApJ...507..558H,
1999ApJ...512...30C,1999AJ....117..102B,2000MNRAS.311..707M}. In
particular, the recent availability of $1024 \times 1024$~pixel IR
detectors has made possible surveys for significant numbers of EROs
\citep{1999ApJ...523..100T}.

The use of the designation ``extremely red'' has varied in the
literature. This was especially true in earlier work where the
description was applied to any object appearing in IR images but
undetected optically. In this paper, we follow
\citet{1996ApJ...471..720G} and use the term ``extremely red object''
(ERO) for a source with $R-K>6$.  This criteria was an operational one,
encompassing several IR-detected, optically-invisible sources known at
the time, and it has since become widely used.  In particular,
\citet{1999ApJ...523..100T}, who have conducted the widest deep $R-K$
survey to date, also adopt this definition. They confirm EROs defined by
this criteria are unusual objects, being the reddest 2\% of the
$K\leq20$ field galaxy population.  However, we will also pay attention
in this paper to galaxies with $R-K>5$, the criteria used by
\citet{1988ApJ...331L..77E} and \citet{1999ApJ...512...30C}, as this is
about the expected color of passively-evolving elliptical galaxies at
$z\gtrsim1$.  We will generically refer to this larger sample as ``red
galaxies.''

EROs are very optically faint ($R\gtrsim24$) which has hampered studies
of these objects.  Successful spectroscopy to determine their redshifts
and physical nature has only become possible with the development of the
Keck 10-m telescope.  Recent work has found some members of this
``missing population'' are luminous ($\gtrsim L^*$) galaxies at $z>1$;
hence determining their origin has direct relevance to the formation of
massive galaxies and AGN.  Being solely defined by optical/IR color, the
few EROs with measured redshifts form a heterogeneous population as
expected, comprising at least two broad classes: (1) ultraluminous dusty
star-forming systems, perhaps akin to local objects like Arp~220, and
(2) massive galaxies with old passively-evolving stellar populations.

The best studied ERO to date, ERO~J164502+4626.4 \citep[object 10
of][hereinafter ``HR10'']{1994AJ....107.1303H}, is the prototype for
star-forming EROs.  This object is a $z=1.44$ dusty, ultraluminous
galaxy with enormous ongoing star formation
($\gtrsim1000$~\Msun~yr\perone) suggested by its sub-mm continuum
emission \citep{1996ApJ...471..720G,1998Natur.392..895C,
1999ApJ...519..610D}.  However, since HR~10 has among the reddest colors
($I-K=6$) of known EROs, it is unrepresentative of the bulk of the
population or at least is an extreme example.  These EROs may provide
excellent case studies of optically obscured star formation in the early
Universe, especially given the association of faint EROs with some of
the sub-mm emitting sources found by SCUBA \citep{sma99}.  In fact, it
may be that intense, very brief bursts of star formation are a common
mode of star formation at high redshift.

There are also examples of EROs as galaxies with old stellar
populations.  Passively evolving ellipticals at $z>1$ are expected to
have large $R-K$ colors, which stem from their rising spectral energy
distributions (SEDs) longward of $4000$~\AA\ being redshifted into the
near-IR.  Therefore, selection using very red colors is an excellent
method to search for early-type galaxies at $z>1$. This is especially
true in searching for clusters since the surface density of galaxies
becomes quite high at faint magnitudes and it would otherwise be hard to
distinguish a cluster from the foreground and background populations.
However, ellipticals at these distances are expected to be optically
very faint so measuring spectroscopic redshifts is challenging.
\citet{dic95} has identified a cluster of elliptical galaxies associated
with the powerful radio galaxy 3C~324 at $z=1.206$ along with a
foreground structure at $z=1.15$.  To date, the highest-redshift
collection of old, red galaxies spectroscopically confirmed has been
found by \citet{1997AJ....114.2232S}. They have discovered a $z=1.27$
cluster by its large $J-K$ colors; the cluster galaxies have
$R-K\gtrsim5$ and rest-frame UV spectra resembling local elliptical
galaxies. There is also a neighboring cluster at $z=1.26$ found from its
X-ray emission by \citet{ros99} which contains spectroscopically old red
galaxies.  Isolated examples of old EROs have been discovered at still
higher redshifts. The very weak radio sources LBDS~53W091 at $z=1.552$
\citep{1996Natur.381..581D,1997ApJ...484..581S} and LDBS~53W069 at
$z=1.432$ \citep{dun98,dey00} both have $R-K\approx6$ and rest-frame UV
spectra which imply ages of a few Gyr.
Recently, \citet{soi99} have identified an $R-K\sim7$ object as an old
galaxy at $z=1.58$ based on associating a large continuum break observed
at $\approx$1~\micron\ with redshifted 4000~\AA\ break.

Discovery of $z>1$ galaxies with old stellar populations offers several
powerful lines of inquiry into understanding galaxy formation and
evolution and its cosmological context. Detailed comparison of the
absorption lines and continuum breaks of these galaxies with galaxies at
$z=0$ may prove fruitful in tracing the evolutionary course and
enrichment history of the oldest stellar populations.  Absolute age
dating of these galaxies would provide a constraint on the time scale of
galaxy formation and the age of the Universe.  Moreover, clusters of old
galaxies at high redshift can provide testing grounds for competing
scenarios of galaxy formation; the predicted appearance of early-type
galaxies in these clusters is dramatically different in hierarchical
galaxy formation scenarios as compared to monolithic collapse ones
\citep[e.g.,][]{1998MNRAS.294..705K}.  Finally, the existence of these
old galaxies at high redshift potentially can constrain cosmological
parameters and theories of structure formation
\citep[e.g.,][]{1998MNRAS.296.1089P}.

We are conducting an on-going study of the nature of these EROs, using
deep optical and near-IR imaging from the ground to assemble a large
sample of EROs for statistical study.  We have been acquiring
high-resolution morphological information from \HST\ optical and Keck
near-IR imaging, and we are obtaining spectroscopy from Keck in the
optical and near-IR to determine ERO redshifts and physical
properties. Deciphering the identity of EROs based on comparing
broad-band colors alone to theoretical stellar population synthesis
models is dubious given that the model parameter space is vast, at least
comprising age, metallicity, and reddening variations. Spectroscopy is
essential.  In addition, given the apparent heterogeneity of the
population, a reasonably large sample of objects needs to be studied to
understand the nature and relative abundances of the subsets, instead of
the spectroscopy of individual EROs which has been done to date.

In this paper, we present a study of the EROs in the field of
QSO~1213--0017 (RA = 12$^h$15$^m$49.8$^s$, Dec = --00\degree 34\arcmin
34\arcsec\ ; J2000.0). This $z=2.69$ quasar, also known as UM~485, has
exceptionally strong and complex Mg~II absorption systems at $z=1.3196$
and $z=1.5534$ \citep{1992ApJS...80....1S}. In \S~2, we describe optical
and near-IR imaging covering an 11~arcmin\pertwo\ region around this
field and Keck optical spectroscopic follow-up of galaxies selected by
their very red colors.  We consider in \S~3 the surface density of the
red galaxies, their morphologies, and their spectroscopic redshifts.  We
examine in \S~4 the collection of red galaxies as a whole, both their
spatial distribution to consider the possibility that they are members
of a cluster at $z=1.31$ and their spectrophotometric properties to
understand their stellar populations.  We summarize our findings in \S~5
and discuss their implications.

Throughout this paper, we assume a cosmology with $\Omega=1$,
$\Lambda=0$, and $H_0=50\ \hf$~\kms~Mpc\perone. At $z=1.31$ with these
parameters, 1\arcsec\ = 8.58~\hfperone~kpc; the luminosity distance $d_L
= 9.40$~\hfperone~Gpc; and the angular diameter distance
$d_\theta=1.78$~\hfperone~Gpc.

%

\section{Observations}

\subsection{Optical Imaging \label{data-optical}}

Optical imaging of the field of QSO~1213--0017 was obtained in April
1993 using the KPNO 4-m Mayall telescope as part of a program to image
$z > 1$ Mg~II absorbing galaxies near QSO sightlines
\citep[see][]{ste95}. We used the PFCCD equipped with a $2048\times2048$
Tektronix CCD and a pixel scale of 0\farcs47/pixel.  A total integration
of 3500~s was obtained through the $\Rs$-band filter
($\lambda_c=6930$~\AA, $\Delta\lambda=1500$~\AA) under photometric
conditions and 1\farcs25~FWHM seeing.  The data were reduced using
standard techniques and calibrated onto the AB magnitude
system\footnote{Note that our preliminary results in \citet{liu99}
designated the \Rs\ filter as ``$R_S$'' and presented Vega-based
magnitudes, instead of AB ones.  For the $\Rs$ filter, the offset from
AB mags to Vega-based mags is --0.28~mag, \ie, $\Rs^{Vega} =
\Rs^{AB}-0.28$.\label{vegamag}} using spectrophotometric standard stars
from the lists of \citet{1988ApJ...328..315M} and
\citet{1983ApJ...266..713O}.  The \Rs\ filter is a compromise between
the standard Cousins $R_C$ and $I_C$ filters.  A filter trace for \Rs\
and a photometric transformation to the Kron-Cousins system are given in
\citet{1993AJ....105.2017S}. For the very red objects considered in this
paper, $\Rs \approx R_{C}$ after accounting for the color terms, so the
$\Rs-K$ colors given for the EROs in this paper are roughly equivalent
to $R_C-K$ colors given for other EROs in the literature (see
\S~\ref{ero-counts}).

We obtained four images of the 1213--0017 field with the WFPC2 camera
aboard {\sl Hubble Space Telescope} (\HST) on U.T.\ 1998 January 11.
Data were taken using the $F814W$ filter with a total integration of
4500~s.  The integrations were taken in slightly offset positions,
allowing reasonable identification of cosmic ray hits in the individual
images which were then excluded from the averaging to create the final
image.

\subsection{Near-IR Imaging}

\subsubsection{KPNO \label{ir-kpno}}

Infrared data were obtained in February 1994 on the KPNO 4m Mayall
telescope using the IRIM imager, equipped with a $256\times256$ NICMOS-3
HgCdTe detector and sampling the sky at 0\farcs60/pixel. A total
integration time of 3660~s was obtained in a non-repetitive grid, in
which the telescope was moved after each 60~s of integration
time. Conditions were photometric with 1\farcs0 FWHM seeing. The data
were reduced using the DIMSUM package (Eisenhardt, Stanford, Dickinson,
\& Ford, priv.\ communication) in IRAF.\footnote{IRAF is distributed by
the National Optical Astronomy Observatories, which are operated by the
Association of Universities for Research in Astronomy, Inc., under
cooperative agreement with the National Science Foundation. } The data
were calibrated using the UKIRT faint standards \citep{cas92}, resulting
in Vega-based magnitudes.  The data were taken in the $K_S$ filter
\citep{1995ApJS...96..117M} which is slightly bluer and narrower than
the standard $K$ filter. As discussed by \citet{1998ApJS..119....1H},
the shift between $K_S$ and $K_{UKIRT}$ is expected to be quite small,
less than 0.04 mag for objects as red as $H-K=1$.  In the absence of IR
colors for all the objects to compute color terms, we assume $K_S=K$
hereinafter, which will have a negligible effect on any of our results.

\subsubsection{Keck \label{data-nirc}}

We obtained deeper, higher spatial resolution IR imaging of two
sub-fields on U.T.\ 1998 May 14--15 using the facility near-IR camera
NIRC \citep{nirc} of the 10-meter W.~M.~Keck~I Telescope located on
Mauna Kea, Hawaii.  The camera employs a Santa Barbara Research
Corporation 256 $\times$ 256 InSb array and has a pixel scale of
0\farcs150/pixel resulting in a 38\arcsec\ field.  We observed 2 fields
close to the quasar, one to the southwest (containing R4, R5, and R6)
and one to the northeast (containing R7, R9, and R10).  The SW field,
centered about (--27\arcsec\ E, --7\arcsec\ N) from QSO~1213--0017, was
observed with the standard $JHK$ filters with integration times of 900,
2040, and 2280~s, respectively, totaled over the two nights. The NE field,
centered about (27\arcsec\ E, 10\arcsec\ N) from QSO~1213--0017, was
observed in $J$ for 780~s and in $K$ for 1200~s on 15~May~1998~UT. We
used integration times of 6 or 20~s per coadd, depending on the
filter. After one minute of integration was coadded, the sum was saved as
an image, and then the telescope was offset by a few arcseconds.  The
telescope was stepped through a non-redundant dither pattern, and an
off-axis CCD camera was used to guide the telescope during the
observations. Conditions were non-photometric while observing the NE
field and also for part of the SW field observations.  For the SW field,
we culled frames which were non-photometric from the reduction
process. For the NE field, the data are used only to examine
morphologies.

The data were reduced in a manner typical for near-IR images.  An
average dark frame was subtracted from each image to remove the bias
level. We constructed a flat field by median averaging scaled images of
the twilight sky. A preliminary sky subtraction of the images was
performed to identify astronomical objects.  Then for each image, we
subtracted a local sky frame constructed from the average of prior and
subsequent images, excluding any of the identified astronomical objects
from the averaging.  We used the brightest source in each field to
register the reduced frames, which were then shifted by integer pixel
offsets and averaged to assemble a mosaic of the field.  Bad pixels were
identified by intercomparing the registered individual images and masked
during the construction of the mosaic.  All these reductions were done
using custom software written for Research System Incorporated's IDL
software package (version 5.1).

We observed the faint HST IR standard SJ~9143
\citep{1998AJ....116.2475P} as a flux calibrator so the resulting
magnitudes are Vega-based.  Aperture photometry of the registration
object in each individual frame verified the SW field data which were
retained were nearly photometric, with any systematic errors of order
5\% or less.  There are no obvious point sources in the images, but the
most compact objects have a FWHM of $\approx0\farcs5$.

\subsection{Optical/Near-IR Photometry}

To identify very red objects, we compiled a photometric catalog of all
objects in the KPNO $K$-band mosaic using the SExtractor software of
\citet{1996A&AS..117..393B}, version 2.0.21.  Each pixel in the $K$-band
image was multiplied by the square root of its exposure time to create a
mosaic with uniform noise over the entire field. Objects in the
noise-normalized mosaic were then identified as any set of contiguous
pixels 1.5$\sigma$ above the background level with an area equal to a
FWHM-diameter circular aperture (10 pixels), \ie, a highly significant
detection, and then aperture photometry was done on the original
$K$-band image.  We used the resulting ``${\tt MAG\_BEST}$'' magnitudes,
which for uncrowded objects use apertures determined from the moments of
each object's light distribution.  This method is similar to that of
\citet{1980ApJS...43..305K} and is designed to recover most of an
object's flux while keeping errors low.  The $\Rs$ image was transformed
to be registered with the $K$ image, and photometry was done using the
same apertures for each object as for the $K$-band image.  The Galactic
reddening towards this field is small, $E(B-V)=0.02$
\citep{1998ApJ...500..525S}, so we apply no extinction corrections to
the photometry.

We ran Monte Carlo simulations to verify the accuracy of the measured
magnitudes and errors.  We inserted 10 artificial stars of known
magnitude and color into the images and then processed the images with
SExtractor in the same fashion as the original data.  For each magnitude
and color bin, we ran the simulation 30 times for a total of 300 stars
and then compared the recovered magnitudes and colors to the input ones
in order to determine the random and systematic photometric errors. The
artificial stars were chosen to span a range in magnitude and color
encompassing all the real objects, with 0.5~mag steps in $K$ magnitude
and 1~mag steps in $\Rs-K$ color.  A low-order polynomial surface was
then fitted to the photometric errors from the simulations as a function
of $K$ magnitude and $\Rs-K$ color, and the fit coefficients were used
to assign photometric errors for the real objects.  In all cases the
systematic offset from the input magnitude to mean recovered magnitude
for the artificial stars was always much smaller than the 1$\sigma$
random errors, so we applied no systematic corrections to the photometry
of the real objects.  The errors produced by SExtractor were found to
agree well with the errors deduced from the simulations, except for
objects near the detection limit. In principal, there will be a surface
brightness dependence since errors measured from stars will not be the
same as those measured from less compact sources such as galaxies
\citep[e.g.,][]{1998ApJ...505...50B}. However, this effect is more
important for determining the completeness corrections, which is not
relevant to our work here, and, furthermore, at the faint limit the
majority of our sources are nearly unresolved so any error in measuring
the error will be negligible.

Since the Keck/NIRC imaging reaches fainter magnitudes than the KPNO
data, they were processed separately, but using SExtractor in an
analogous fashion.  Objects were identified and photometric apertures
were determined from the $K$-band image. This information was used to
process the other bands, with Monte Carlo simulations used to compute
photometric errors.

\subsection{Optical Spectroscopy}

We observed the 1213--0017 field with the 10-m Keck~II Telescope using
the facility optical spectrograph LRIS \citep{1995PASP..107..375O} on
U.T.\ 1999 June 15.  We prepared a focal plane mask of 1\farcs5-wide
slitlets to observe seven objects with $\Rs-K>5$.  The length of each
slitlet varied, with a minimum length of 14\arcsec. We used the 150
lines/mm grating blazed at 7500~\AA\ to cover a nominal wavelength
region from 4000--10000~\AA, depending on the position of the slitlets
on the mask. The dispersion of $\approx$4.8~\AA/pixel resulted in a
spectral resolution of $\approx$25~\AA, as measured by the FWHM of a
few strong emission lines of serendipitous objects and the lines of the
calibration arc lamp spectra.  Sky conditions were photometric with
seeing of 0\farcs45 at the start of the integrations.  We obtained 3
exposures of 1800~s each with the mask oriented at a position angle of
46\degree\ east of north. We dithered the telescope a few arcseconds
along the slit direction between each exposure.

The slitmask data for the objects were separated into seven individual
spectra and then reduced using standard longslit tasks in IRAF. The data
were bias-corrected using the unilluminated pre- and post-scan regions
of the detector and flat-fielded using internal quartz lamps obtained
immediately after the observations.  Sky lines were subtracted by
fitting a low-order analytic function to each column of the images,
perpendicular to the dispersion axis. We corrected each exposure for
long-wavelength fringing by taking an average of the other 2 exposures
and subtracting this average.  The exposures were registered and
combined.  The most compact objects in the resulting mosaic are
$\approx0\farcs6$~FWHM in the spatial direction, though they may not be
point sources. A 1-d spectrum of each red galaxy was extracted using a
1\farcs3 (6 pixel) wide aperture and was wavelength-calibrated using
HgKr and NeAr lamps taken immediately after the observations, with an
RMS of $\approx0.3-0.4$~\AA\ in the wavelength solutions. Zeropoint
shifts of $\leq2$~\AA\ were applied to the spectra based on the measured
wavelengths of the night sky lines. Flux calibration was performed using
the standard star G191B2B
\citep{1988ApJ...328..315M,1990ApJ...358..344M}


\section{Results}

\subsection{An Excess of EROs \label{ero-counts}}

\begin{figure}
\centering
\vskip -0.3in
\hbox{
\hskip -1.3in
\includegraphics[angle=180,width=6in]{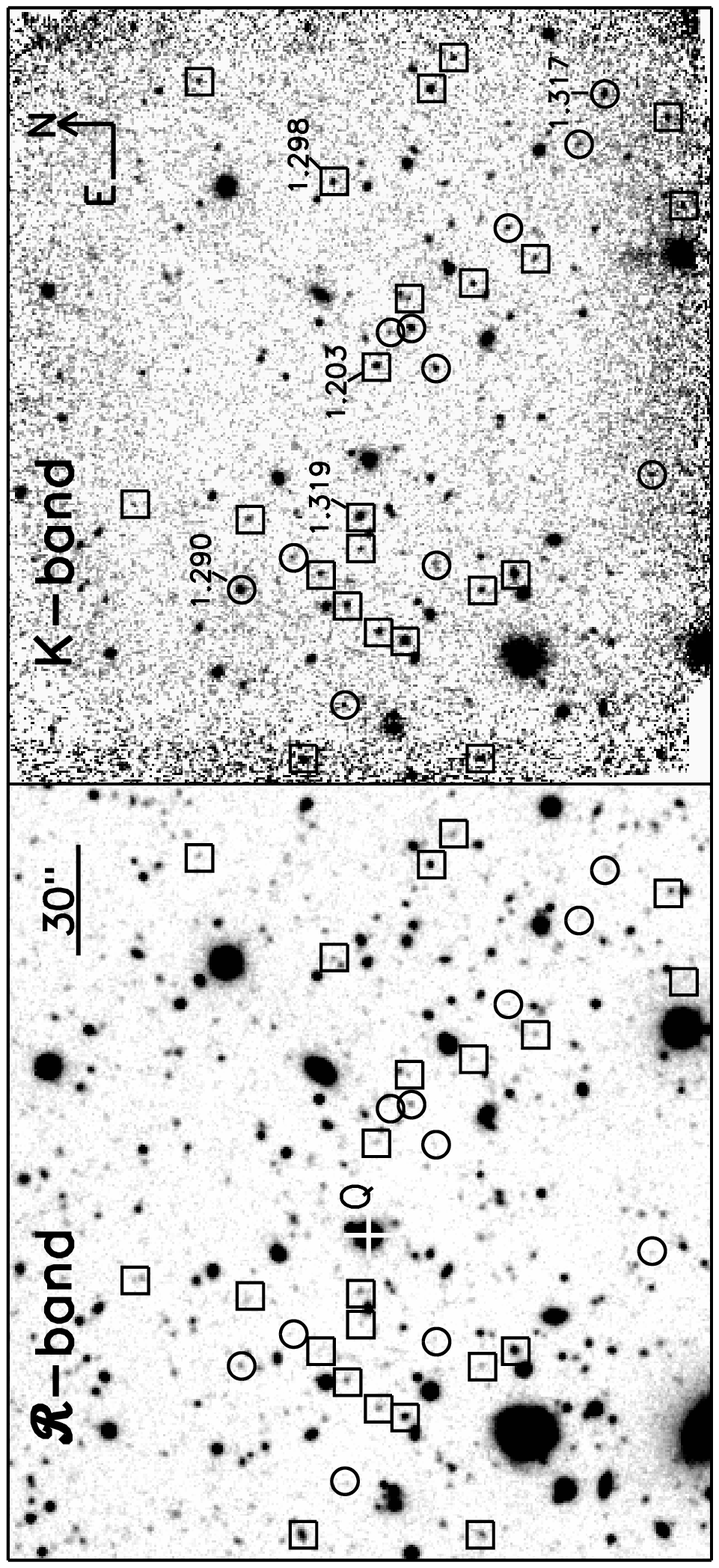}}
\caption{KPNO $\Rs$ and $K$-band images of the 1213--0017 field. Each is
3\farcm5 $\times$ 3\farcm1. Objects with $\Rs-K>6$ are circled
($\bigcirc$) and those with $\Rs-K=5-6$ are marked with a square
($\Box$).  The quasar is marked on the $\Rs$-band image with a white
cross and labelled ``Q''. Red galaxies with spectroscopic redshifts are
labelled. \label{twopanel}}
\end{figure}

\begin{figure}
\hbox{\hskip -0.5in
\centering \includegraphics[angle=90,width=7in]{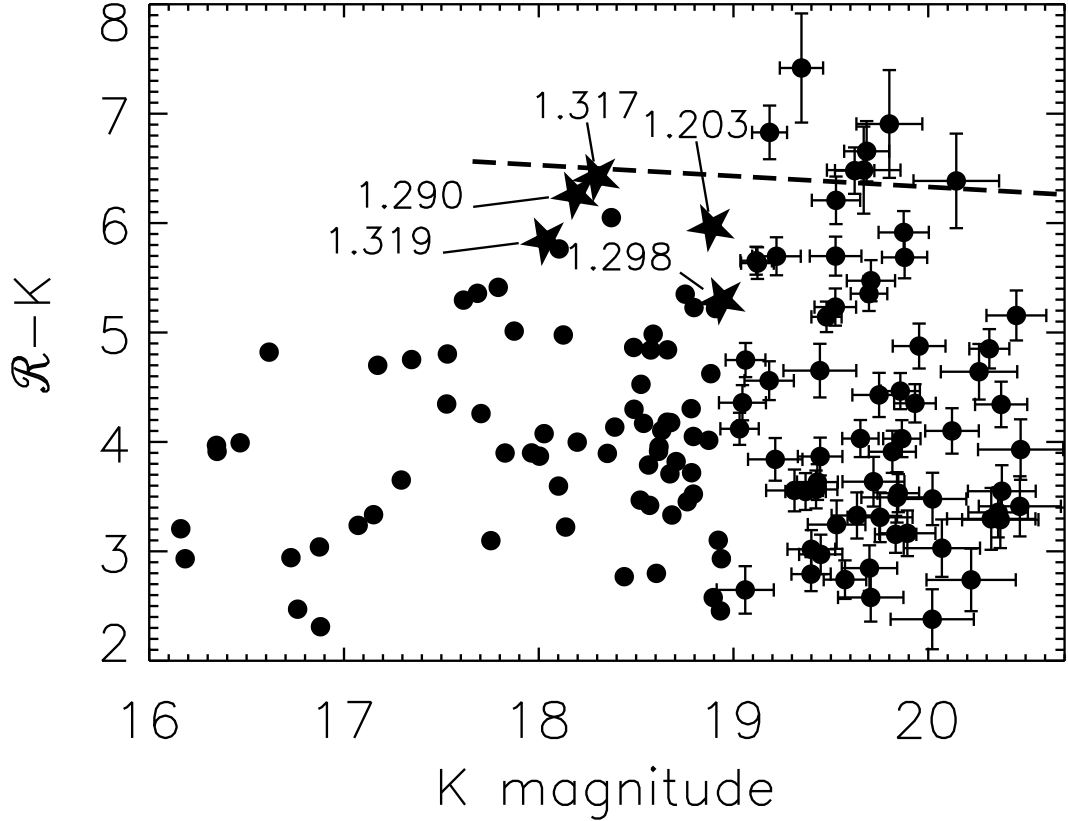}}
\caption{Color-magnitude diagram for objects detected in the
11~arcmin$^2$ $K$-band image shown in Fig. \ref{twopanel}. The error
bars for $K<19$ objects are smaller than the plotting
symbols. $\Rs$-band magnitudes are AB-based, and $K$-band mags are
Vega-based.  There are 11 objects with $\Rs-K>6$ and and another 22 with
$\Rs-K=5-6$. The dashed line represents a no-evolution color-magnitude
locus for elliptical galaxies at $z=1.3$ based on Coma cluster galaxies
(Stanford \etal\ 1998), with the bright end indicating the expected
location of the brightest cluster galaxy. The stars ($\star$) show the
five objects with spectroscopic redshifts. \label{cmd}}
\end{figure}

Figure~\ref{twopanel} presents the KPNO $\Rs$-band and $K$-band images
of the 1213--0017 field, which cover an area of 11.2~arcmin\pertwo.
Figure~\ref{cmd} shows the field's color-magnitude diagram along with
the no-evolution locus of local elliptical galaxies redshifted to
$z=1.3$, computed from multicolor photometry of the Coma cluster as
described in \citet{1998ApJ...492..461S}.  If we were to account for
passive stellar evolution, the locus would appear somewhat bluer and
brighter.

A population of objects with red ($\Rs-K>5$) colors are found starting
at $K\approx18$~mag, irregularly distributed on the sky.  Most of the
galaxies are somewhat bluer than the Coma locus at $z=1.3$, and the
brightest of the EROs have $K$-band fluxes comparable to the brightest
Coma cD galaxies.  As we discuss below, there is a significant excess of
red objects in this field compared to the surface density measured from
blank sky surveys.
However, before analyzing the counts of the reddest objects, we must
first account for a few issues: (1) the effective wavelength of the
$\Rs$ filter is significantly redder ($\approx350$~\AA) than the
standard Cousins $R_C$ filter; (2) our data becomes incomplete at
$K\gtrsim19.5$, as gauged by the turnover in the number counts; and (3)
the exposure time is not uniform over the field, with only about half
the field receiving at least 50\% of the maximum integration.  (Also,
there will be some added uncertainty since the abundance of the reddest
objects has a strong dependence on the choice of color and magnitude
cuts.)

The most extensive $R-K$ color-selected counts in a blank field of the
sky come from the CADIS survey of \citet{1999ApJ...523..100T}.  They
find a surface density of $0.039 \pm 0.016$~arcmin\pertwo\ in an area of
154~arcmin$^2$ for the bright EROs ($R-K\arcmin > 6$ and $K\arcmin <
19$), slightly revised to $0.047 \pm 0.009$~arcmin\pertwo\ in their
final survey area of 600~arcmin$^2$ (D.\ Thompson, priv.\
communication).  Direct comparison of our ERO counts with theirs is
complicated as their $R_{CADIS}$ filter is $\approx450$~\AA\ bluer than
our $\Rs$ filter.  Because EROs are very red and their number counts are
a strong function of $R-K$ color, the difference is significant.  (The
difference in IR filters, their $K\arcmin$ versus our $K_S$, is much
less important.)  We computed the offset between $R$ filters as a
function of redshift using the elliptical galaxy SED of
\citet{1980ApJS...43..393C}.  The true SEDs of the ERO population(s) are
unknown, but we adopt this template as a reasonable guess.  At $z=0$,
$(R_{CADIS}-\Rs)$ is --0.08~mag; the difference is negative due to the
fact that \Rs\ mags are AB and $R_{CADIS}$ mags are Vega-based (see
footnote~\ref{vegamag}). This difference increases nearly monotonically
with redshift out to $z\approx1.5$, becoming as large as 0.4~mag.  For
redshifts of $0.5-1.7$, the calculated offset is $\ge0.17$~mag.  Since
the few EROs with known redshifts lie in this range, we add 0.17~mag to
our measured $\Rs-K$ colors to compare to the Thompson~\etal\
results. This is the minimum expected amount for the transformation
(\ie, an ERO at $z=0.5$) and therefore leads to conservative estimates
for the 1213--0017 ERO overdensity.

After accounting for this difference in filters, we find five $K\le19$
objects with $(R_{CADIS}-K)>6$ in our images, four of which lie in the
central 6.2~arcmin$^2$ area which received least half of the maximum
integration time.  Since the counts over the whole mosaic are guaranteed
to be incomplete, we consider only this central region. The counts of
\citet{1999ApJ...523..100T} predict 0.24~objects in a 6.2~arcmin$^2$
area, 16$\times$ fewer than observed. The Poisson probability of our
observing four or more EROs given the expected surface density is only
$1.1 \times 10^{-4}$, \ie, the overabundance of EROs in the 1213--0017
field is highly statistically significant.  The probability grows to 5\%
if the blank sky counts are increased by a factor of 5.7, \ie, the
1213--0017 ERO surface density is overabundant by this factor at the
95\%~confidence level. 

Similarly, with the less restrictive color criteria of $(R_{CADIS}-K) =
5-6$, where the blue cutoff is about the color of a passively evolving
single-burst population at $z\gtrsim1$ (Figure~\ref{ero-colors}), the
\citet{1999ApJ...523..100T} data predict 2.4~objects with $K\le19$ while
we observe 7~objects. The Poisson probability of finding at least this
many objects is $1.2\times10^{-3}$, and the enhancement of such objects
in the 1213--0017 field is at least a factor of~1.4 at the
95\%~confidence level.

We can combine the counts in the $(R_{CADIS}-K)>6$ and
$(R_{CADIS}-K)=5-6$ bins to compute the enhancement of the 1213--0017
field over the blank sky counts.  We compute the joint probability of
observing the 4 and 7 galaxies in these two color bins over a range of
enhancement factors and find the most likely excess is a factor of 4.2,
with the 95\% confidence range being factors of 2.2--7.1. Note that this
calculation assumes the enhancement is the same for both color bins,
which may not be the case.

\subsection{Morphologies of the Red Galaxies}

\begin{figure}
\centering
\hbox{
\includegraphics[angle=90,height=6in]{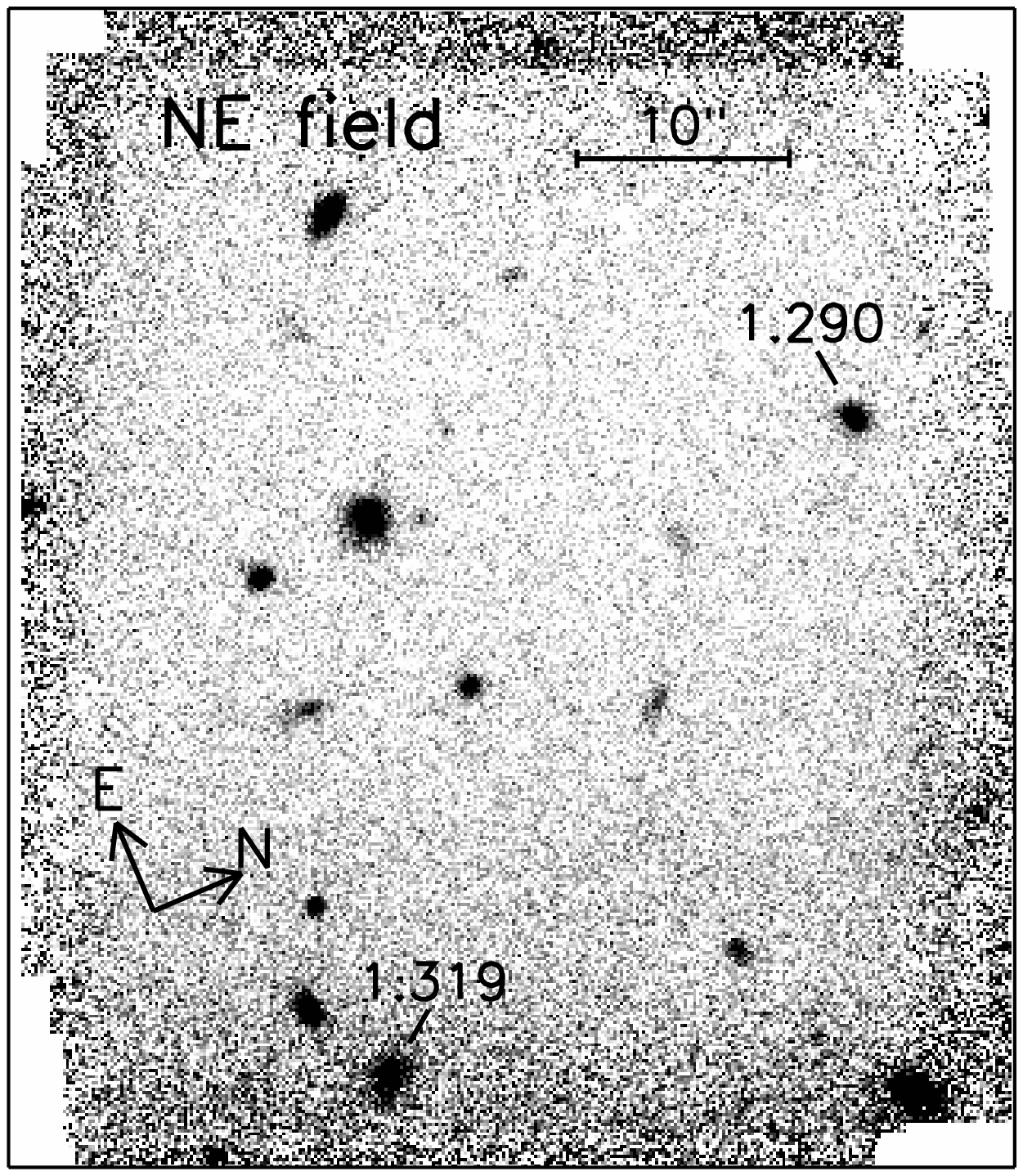}}
\caption{Keck $K$-band image of a field NE of QSO~1213-0017 (see
\S~\ref{data-nirc} for position) shown with a linear stretch. The field
is 47\arcsec\ $\times$ 55\arcsec. The very faintest objects in the
images are $K\approx20.5$~mag. EROs with spectroscopic redshifts are
labeled. \label{nirc-ne}}
\end{figure}

\begin{figure}
\centering
\hbox{
\hskip -0.01in
\includegraphics[angle=90,height=6in]{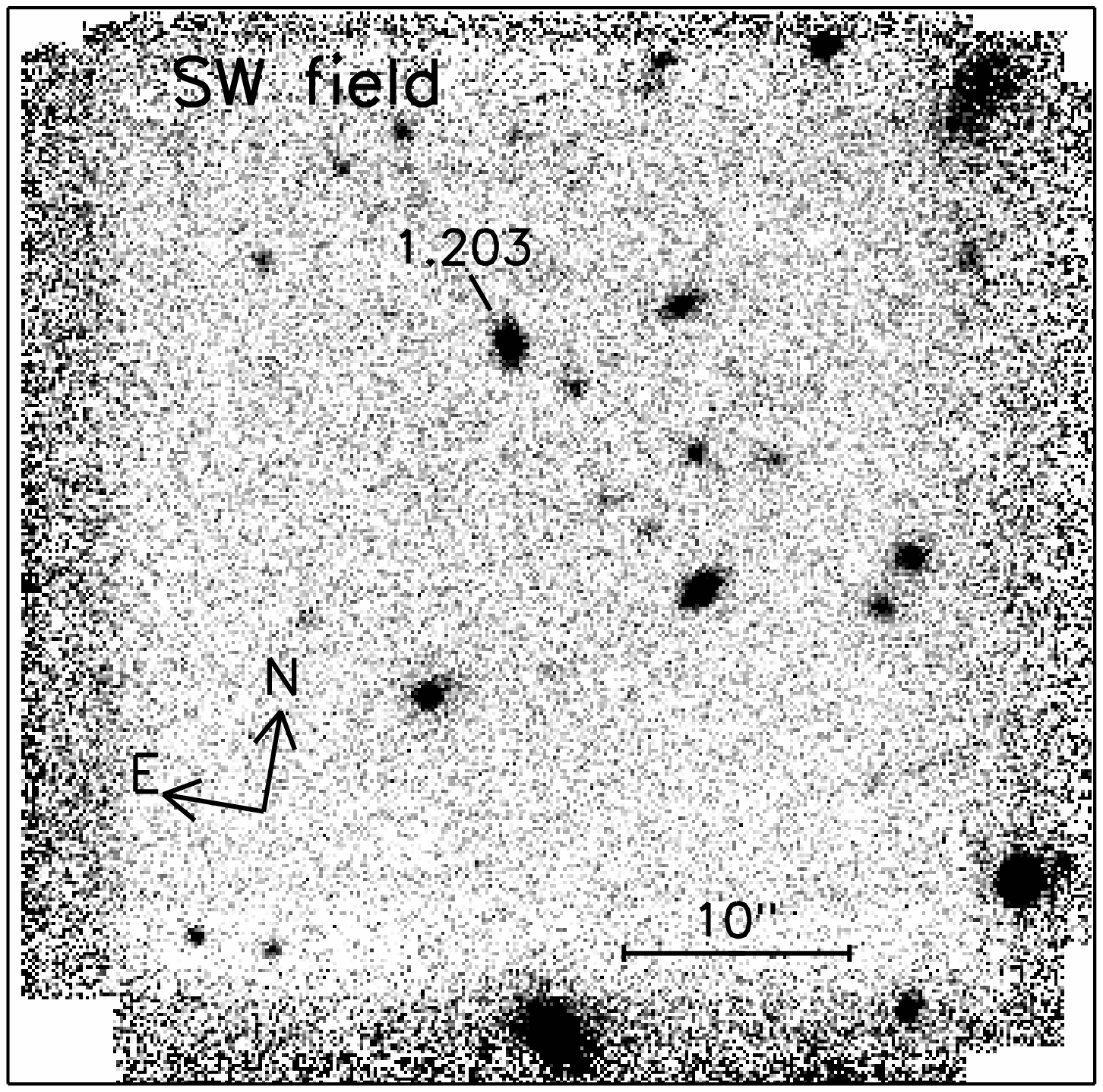}}
\caption{Keck $K$-band image of a field SW of QSO~1213-0017 (see
\S~\ref{data-nirc} for position) shown with a linear stretch. The field
is 48\arcsec\ $\times$ 48\arcsec. The very faintest objects in the
images are $K\approx21.3$~mag. The one EROs in the image with a
spectroscopic redshift is labeled. \label{nirc-sw}}
\end{figure}

Figure \ref{nirc-ne} and \ref{nirc-sw} show the reduced Keck/NIRC
$K$-band mosaics for the fields NE and SW of QSO~1213--0017. Note that
the images reach different limiting depths; the faintest objects in the
NE and SW data are $\approx$20.5 and 21.3~mag, respectively.  We compare
these data with the \HST\ $F814W$ images in Figure~\ref{zoom-contours} to
examine the observed optical and IR morphologies of the objects with
$\Rs-K>5$.  The \HST\ images are centered on the QSO so we do not have
images for the red galaxies farther away, including unfortunately R1 and
R8 for which we have continuum-break spectroscopic redshifts
(\S~\ref{contbreak-spectra}).  For $z=1.3$, these datasets roughly
correspond to the rest-frame near-UV/blue (3100~--~4100~\AA) and far-red
(8700~--~10400~\AA).  \oii\ emission lies in the $F814W$ filter bandpass
for objects at $z\approx0.9-1.5$.  All the red objects are extended in
the \HST\ images meaning they cannot be galactic M~dwarfs.  A few of
them appear to be E/S0 type. The \HST\ images are relatively shallow
($3\sigma$ limiting surface brightness
$\mu_{F814W}\approx25.7$~mag~arcsec\pertwo), but we can still see faint
emission around a number of the red galaxies. Several seem to have both
a central spheroid and a faint component suggestive of a disk, while a
few are entirely diffuse emission without any compact core/bulge. None
show any strong evidence for interactions. Overall, the detected red
galaxies show remarkable consistency in their morphologies between the
$F814W$ and $K$-band images.  We return to this issue in
\S~\ref{red-colors}.

\subsection{Optical Spectroscopy \label{spectra}}

From the 2-d reduced spectra, we were able to identify spectroscopic
features in five of the red galaxies on our slitmask, both from emission
lines and continuum breaks (Table \ref{table-spectra}).  To more
accurately determine the redshifts and their errors, we used the Fourier
cross-correlation technique of \citet{1979AJ.....84.1511T} as
implemented in the FXCOR task of IRAF.  We linearly interpolated over
significant residuals in the spectra resulting from imperfect
cancellation of the strongest telluric features.  The three
absorption-break galaxies were cross-correlated against the UV spectra
of F~and G~dwarfs from the \IUE\ Spectra Atlas of \citet{wu91}, using an
average spectrum for each spectral type.  The spectra of the
emission-line galaxies were cross-correlated against the emission-line
template of an Sc~galaxy from \citet{1996ApJ...467...38K}.

\subsubsection{R6 and R7 --- Emission-Line Galaxies} 

\begin{figure}
\centering
\hbox{\hskip 0.5in
\includegraphics[angle=0,width=5in]{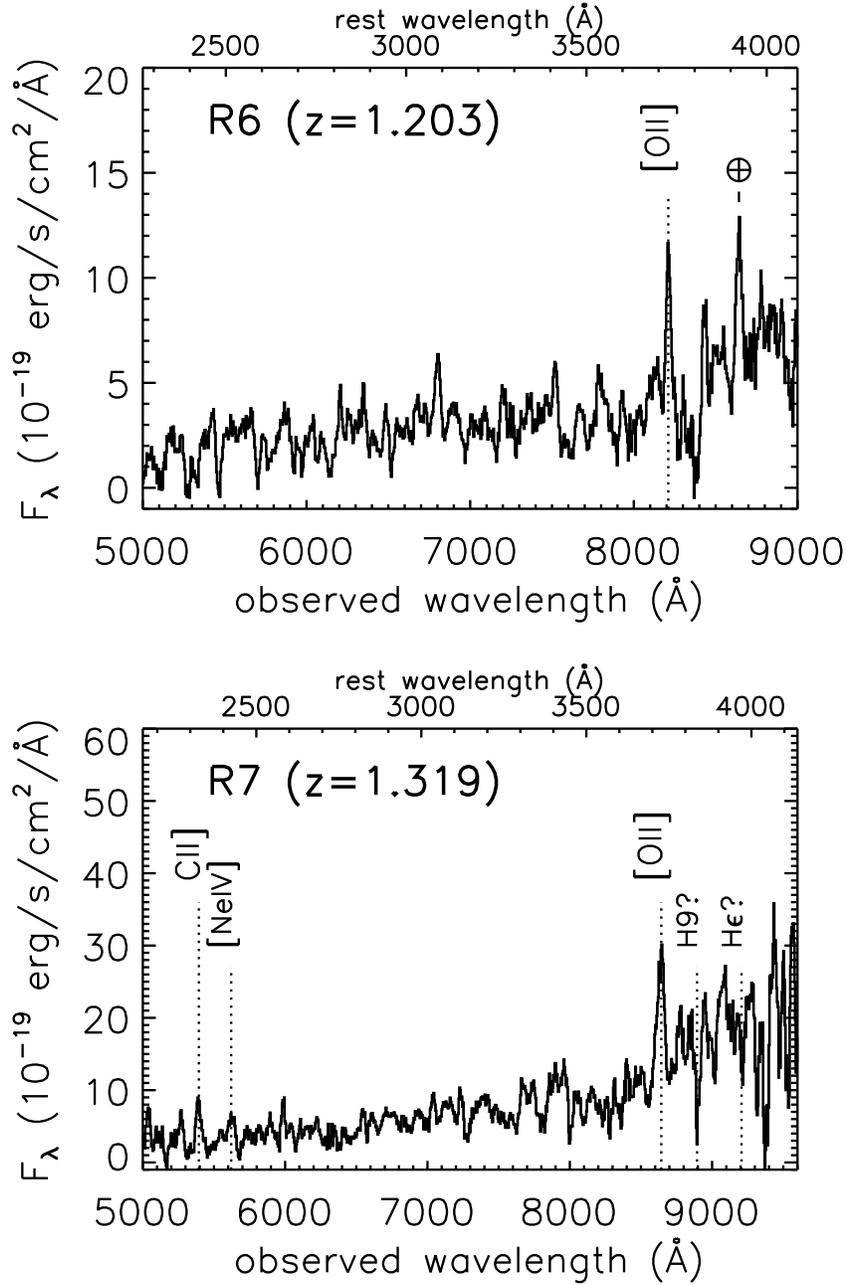}}
\caption{Keck/LRIS spectrum of two emission-line red galaxies.  Both
spectra have been smoothed with a boxcar filter of width 7 pixels. {\bf
Top:} R6 shows a single strong emission line at 8210~\AA, which
we assume is \oii.  The feature at 8642~\AA\ is due to imperfect
cancellation of a CCD fringe from a strong sky line. {\bf Bottom:} R7
shows emission lines of C~II]~$\lambda$2326 and \oii, and also a weak
[\ion{Ne}{4}]~$\lambda$2424 line close to the telluric [\ion{O}{1}] line
at 5577~\AA.  \label{spec-r6+r7}}
\end{figure}

The spectrum of R6 shows a single emission line with an observed
equivalent width (EW) of $\approx$60~\AA\ (Figure~\ref{spec-r6+r7}).
There is ample continuum blueward of the line so we identify it as \oii,
placing R6 at $z=1.203$, slightly in the foreground of the other
redshifts in this field.  Assuming this emission originates from \hii\
regions of massive young stars, we infer a star formation rate of
$\approx10~h_{50}^2$~\Msun~yr\perone\ using the \oiin\ calibration of
\citet{1992ApJ...388..310K}, comparable to the most active spirals in
the local neighborhood \citep{1983ApJ...272...54K}.

The spectrum of R7 reveals the presence of an active galactic nucleus at
$z=1.319$ (Figure~\ref{spec-r6+r7}). The data show strong lines of \oii\
and \ion{C}{2}]~$\lambda$2326 with observed equivalent widths of
$\approx50$~\AA.  The \ion{C}{3}]~$\lambda$1909 line is possibly
detected, though this line falls near the blue end of the spectrum where
the signal is quite low.  Examination of the reduced images clearly
shows a weak [\ion{Ne}{4}]~$\lambda$2424 emission line which lies on the
red edge of the strong [\ion{O}{1}]~$\lambda$6300 telluric line,
preventing a good measurement. Strangely, no \mgii\ emission line is
seen, even though it is typically at least as strong as
[\ion{Ne}{4}]~$\lambda$2424 and \ion{C}{2}]~$\lambda$2326 in high-$z$
AGN \citep[e.g.,][]{1999AJ....117.1122S}. Presumably the emission line
is absorbed by gas in this galaxy though no strong \mgii\ absorption
feature is seen. Also, there are some possible weak absorption features
in the 8800--9200~\AA\ range near the expected location of the higher
order Balmer lines, in particular at 8896~\AA\ which matches the Balmer
H9 transition. However, spectra become quite noisy in this region and
the features are not overly compelling. Finally, these are no signs of
an old stellar population as the strongest continuum breaks (B2640,
B2900, D4000) and the Ca~H+K lines are all absent.

\subsubsection{R1, R8, and R10 --- Old Continuum-Break Galaxies 
\label{contbreak-spectra}}

\begin{figure}
\centering
{\hbox{\hskip 0.5in
\includegraphics[angle=0,width=5.5in]{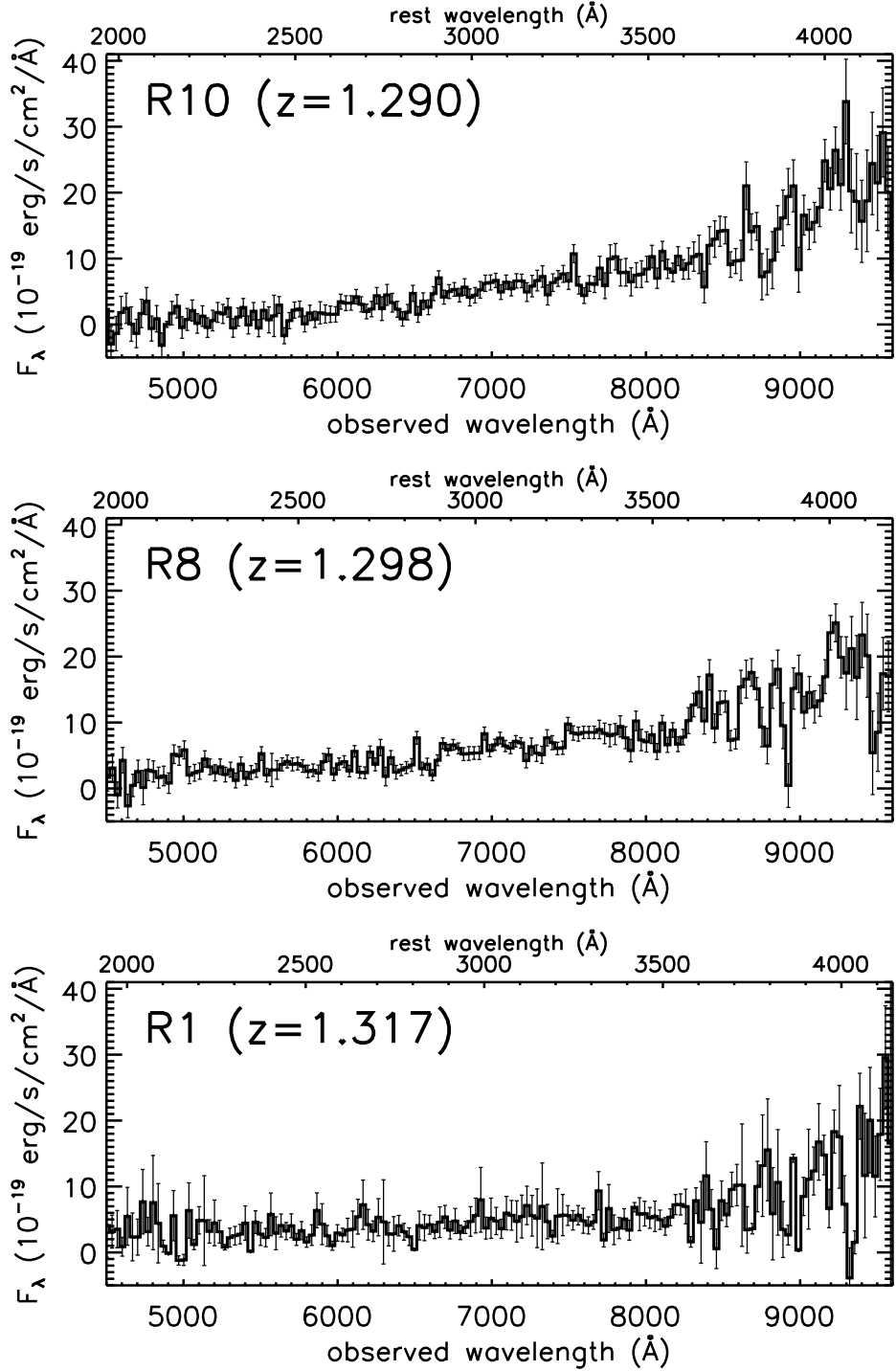}}}
\vskip 0.2in 
\caption{Keck/LRIS spectrum of the three absorption-line red galaxies.
The spectra have been averaged in 7 pixel bins with the formal 1$\sigma$
errors overlaid. See the text and later figures for a description of the
redshift determination. At $\lambda\gtrsim8500$~\AA, the noise of the
spectra increases due to telluric OH emission lines and also imperfect
removal of CCD fringes. \label{errspec}}
\end{figure}

Three of the red galaxies on our slitmask --- R1, R8, and R10 --- are
devoid of strong emission lines (Figures~\ref{errspec}), but they do
show continuum breaks identifiable as the rest-frame mid-UV
(2000~--~3300~\AA) breaks of B2640 and B2900 at $z=$~1.317, 1.298, and
1.290, respectively. The breaks arise from metal-line blanketing on
either side of the tophat-shaped feature from 2640--2720~\AA.  These
features, which were first suggested by \citet{1977ApJ...212..438M} as
possibly being useful for redshift determinations at $z>0.75$, have been
used recently to identify old galaxies out to $z\approx1.55$
\citep{1996Natur.381..581D,dey00}.  At least two of the red galaxies may
also have \mgii\ in absorption. These spectral features are
characteristic of old stellar populations, such as Galactic and M31
globular clusters, M32, and the cores of local elliptical galaxies
\citep[e.g.,][and references
therein]{1999AJ....117.2213R,1998AJ....116.2297P}.  Galaxies R8 and R10
also may show the 4000~\AA\ break at $\approx9200$~\AA.

\begin{figure}
\centering \includegraphics[angle=0,height=7in]{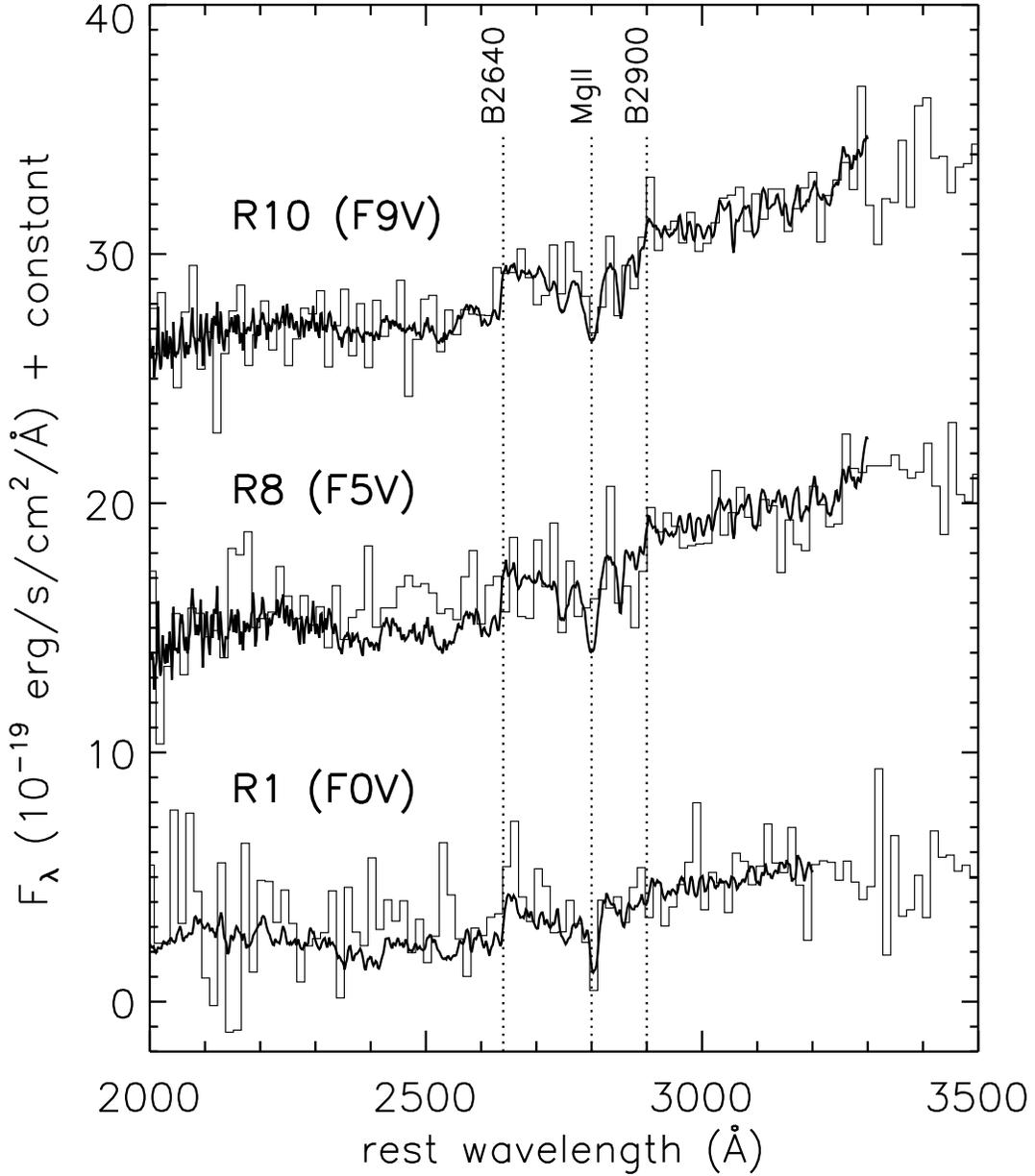}
\caption{Rest-frame UV spectra of the three continuum-break galaxies,
averaged in 7 pixel bins.  The R8 spectrum is offset by +13 along the
y-axis and the R10 spectrum by +26.  Overplotted with the thick lines
are \IUE\ spectra of dwarfs.  The \IUE\ spectra have been normalized to
the galaxy flux longward of 3000~\AA; notice the good agreement between
the star and galaxy spectra all the way to the bluest wavelengths. This
suggests the amount of any ongoing or recent star formation, which would
be revealed by a rising blue continuum, is small.
\label{absrestframe}}
\end{figure}

For these three galaxies, we determined the redshifts using the
4300--8000~\AA\ region of the spectra in the cross-correlation
analysis. At longer wavelengths, the random errors increase from
enhanced telluric OH emission, and the systematic errors also increase
from imperfect removal of the CCD fringing pattern, due to the small
number of exposures we were able to acquire.  In addition to the choice
of wavelength region, another important aspect of the redshift
determination is the choice of stellar template.  We explored template
spectra ranging from spectral types F0 to G8, using both the Fourier
analysis and a direct computation of the cross-correlation between
galaxy and template as a function of redshift. All templates produced
fairly consistent redshifts using both methods.  Figure
\ref{absrestframe} shows the rest-frame galaxy spectra with \IUE\
stellar spectra overplotted for comparison.


\section{Discussion}

\subsection{A Concentration of Red Galaxies at $z$ = 1.31
\label{concentration}}

We have measured five redshifts for the red galaxies in the 1213--0017
field.  Four of these lie near $z=1.31$ with the remaining one at
$z=1.203$.  The spectrum of QSO~1213--0017 also shows an \ion{Mg}{2}
absorption system at $z=1.3196$
\citep{1987ApJ...322..739L,1992ApJS...80....1S}, similar to the redshift
of $1.319\pm0.002$ for galaxy R7. Could this galaxy be responsible for
the absorption?  It is located 15\arcsec\ from the QSO, which is
130~\hfperone~kpc.  Based on preliminary results from a survey of
$z=0.3-0.9$ \ion{Mg}{2} absorbers by \citet{ste93} and \citet{ste94},
the median projected impact parameter between QSOs and their best
candidate absorber is 45~\hfperone~kpc with an upper bound of
80~\hfperone~kpc.  Early results from their $z>1$ survey find the
absorber properties are not dramatically different compared to the lower
redshift sample \citep{ste95,1996IAUS..171..295D}.  The projected
separation of R7 therefore seems too large for it to be the \mgiin\
absorber, though the geometry and extent of \ion{Mg}{2} absorption
systems remain open questions
\citep[c.f.][]{ste95,1996ApJ...465..631C}. In addition, our \HST\ images
show a few faint galaxies with projected separations from the QSO of
$\lesssim80\ \hfperone$~kpc, making them better candidates for the
absorber.  Since R7 is unlikely to be the \ion{Mg}{2} absorber, we infer
that there is at least one more galaxy at $z=1.3196$.

Including the \ion{Mg}{2} absorber, there are five galaxies within an
angular region of 3\arcmin\ and close together in velocity. Their
redshifts have an unweighted mean of 1.309 and a standard deviation
$\sigma=1790\pm570$~\kms\ in the mean rest frame.\footnote{The above
results use the standard formulae.  Similar results are obtained using
the central location (``mean'') and scale (``dispersion'') estimators
recommended by \citet{1990AJ....100...32B} for non-Gaussian velocity
distributions.  Their preferred central location measures for a sample
of this size, the median and the biweight-estimated mean, give $z=1.317$
and 1.319, respectively. The gapper method they advocate for computing
the scale gives 1840~\kms.}  The full spread in redshift is 3800~\kms\
in the mean rest frame; for comparison, consider that 95\% of the
galaxies in the central 6~$\hfperone$~Mpc of the Coma cluster and with
$cz<12,000$~kms lie within $\pm 3500$~\kms\ \citep{1998AJ....116..560P}.
The considerable dispersion and range of the 1213--0017 redshifts are
possibly due to velocity sub-structure in this field.  There is a hint
of this in the data: two of the galaxies (R8 and R10) lie at
$z=1.290-1.298$ while the other three are at $z=1.317-1.320$, \ie, there
is a ``gap'' of 2500~\kms\ in the mean rest frame, although there is no
clear segregation of the two subsets on the sky. We may be seeing two
separate physical entities close in redshift (\eg,
filaments/sheets/sub-clusters) which may later merge into one, but this
speculation lies beyond the available data.  

Interestingly, the few known $z\approx1$ massive clusters show large
velocity dispersions and/or signs of sub-structure, \eg, the 3C~324
``cluster'' at $z=1.15$ and 1.21 \citep{1997gsr..proc..215D};
RX~J1716.6+6708 at $z=0.81$ \citep{1997AJ....114.1293H,
1999AJ....117.2608G, 1998ApJ...497L..61C}; and CL~0023+0423 at $z=0.84$
\citep{1998AJ....116..560P, 1998AJ....116..643L}.  Spectroscopically
confirmed galaxies in these clusters have angular extents of a few
arcminutes but are not strongly concentrated on the sky, which is also
the case for the red galaxies in the 1213--0017 field. In the case of
ClG~J0848+4453 at $z=1.27$ and the neighboring RXJ0848.9+4452 at
$z=1.26$ \citep{1997AJ....114.2232S,ros99}, each individual cluster is
fairly compact but taken together these two add up to an entity with a
large velocity range separated by only a few arcminutes.


In addition to the galaxies with spectroscopic redshifts, we can use
optical/IR colors to constrain the redshifts of objects in the SW field,
the only one with photometric data in all filters (see \S~2).
The galaxies' $\Rs JK$ colors are consistent with Bruzual \& Charlot
(1996) models of old stellar populations at $z=1-2$
(Figure~\ref{ero-colors}), with the identification largely based on the
change in $\Rs-J$ color. The $JHK$ colors provide less sensitivity, but
they are consistent with the same redshift interval
(Figure~\ref{ero-colors-jhk-sw}).


To summarize, the collection of red galaxies in the 1213--0017 field is
concentrated in redshift at $z\approx1.31$.  The small number of
measured redshifts prevents any definitive conclusions, but the
redshifts do have a considerable spread.  The red galaxies are spread
over $\approx 1.5\ \hfperone$~Mpc on the sky, with a surface density
about a factor of four above $R-K>5$ counts and a factor of 16 above
bright ERO counts in blank fields (\S~\ref{ero-counts}).  Although we do
not know all their redshifts, it is quite plausible that most of the red
galaxies are at $z=1.31$, as we discuss in the next section.

\subsection{Old Stellar Populations at $z$ = 1.31}

\subsubsection{Origin of the Extremely Red Galaxy Colors \label{red-colors}}

\begin{figure}
\vskip -1.2in
\hskip -0.3in
\centering \includegraphics[angle=90,width=6in]{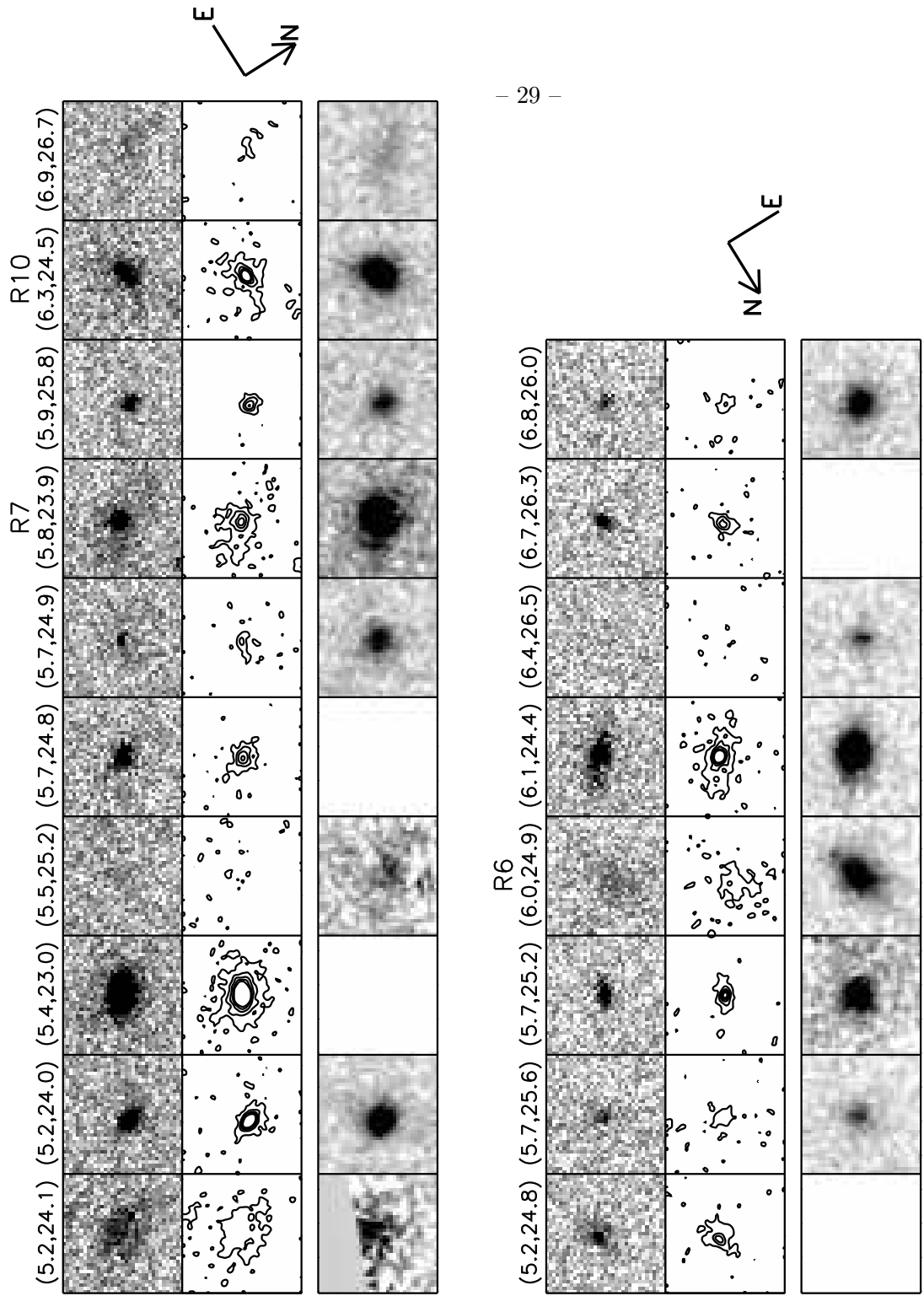}
\vskip 0.1in
\caption[zoom-stamps.ps]{\HST\ $F814W$ and Keck $K$-band images for most
of the red galaxies in the field of QSO~1213--0017. The top set is from
WF3, which covers the field NE of the quasar, and the bottom set from
WF4, covering the SW field.  For each set, the top row shows the
greyscale \HST\ image, the middle row shows contours for the \HST\ data
drawn at $\{\frac{1}{4}, \frac{3}{4}, \frac{5}{4}, \frac{7}{4},
\frac{9}{4}\} \times \sigma$, and the bottom row shows the Keck image,
when available. The orientation on the sky is shown at the right edge of
each set. The images are 3\farcs5 (30~\hfperone~kpc) on a side and
displayed with a linear stretch. The $\Rs-K$ color and \Rs\ mag measured
from our KPNO data are above each object. The three with redshifts (R6,
R7, and R10) are labeled.  Most of the red galaxies have a regular
appearance, and none are obviously interacting or disturbed.
\label{zoom-contours}}
\end{figure}

\begin{figure}
\centering
\hbox{
\hskip -0.5in \includegraphics[angle=90,width=7in]{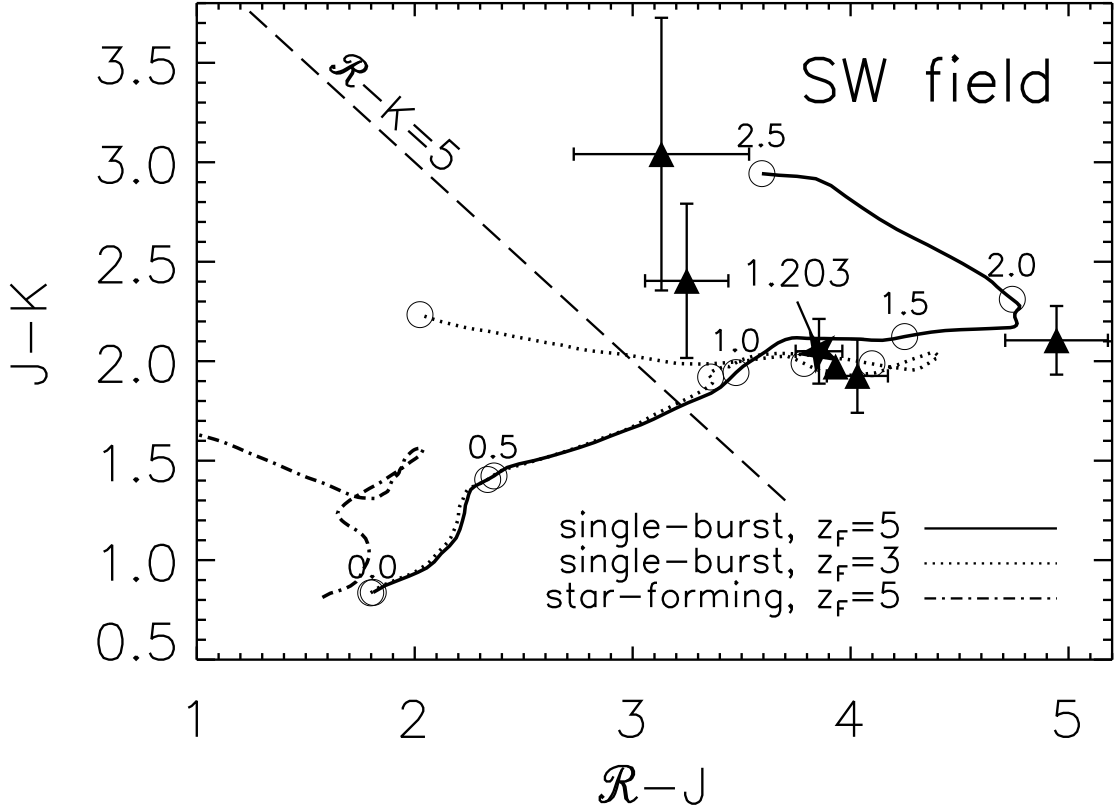}}
\caption{$\Rs JK$ color-color plot of red ($\Rs-K>5$) objects in the
Keck/NIRC field SW of the quasar, the only field with IR photometry in
two filters. The data point without error bars has errors comparable to
its symbol size. Overplotted are single-burst models of Bruzual \&
Charlot (1996) passively evolving from formation redshifts of
$z_F=3$~and~5. The open circles ($\circ$) show the observed model colors
at $z=0-2.5$. The colors of the single-burst models diverge for
$z\gtrsim1.5$.  We also plot colors for an exponentially decaying
star-forming model with an $e$-folding time of 6 Gyr, which well
reproduces the colors of local spiral galaxies (\eg, Charlot \& Bruzual
1991); its colors never become significantly red. The observed colors of
the well-detected red galaxies suggest these objects lie at $z=1-2$,
corroborating the color-magnitude plot in Figure \ref{cmd}. The object
marked with a star ($\star$) is R6, an emission-line ERO at
$z=1.203$. \label{ero-colors}}
\end{figure}

\begin{figure}
\centering
\hbox{
\hskip -0.5in
\includegraphics[angle=90,width=7in]{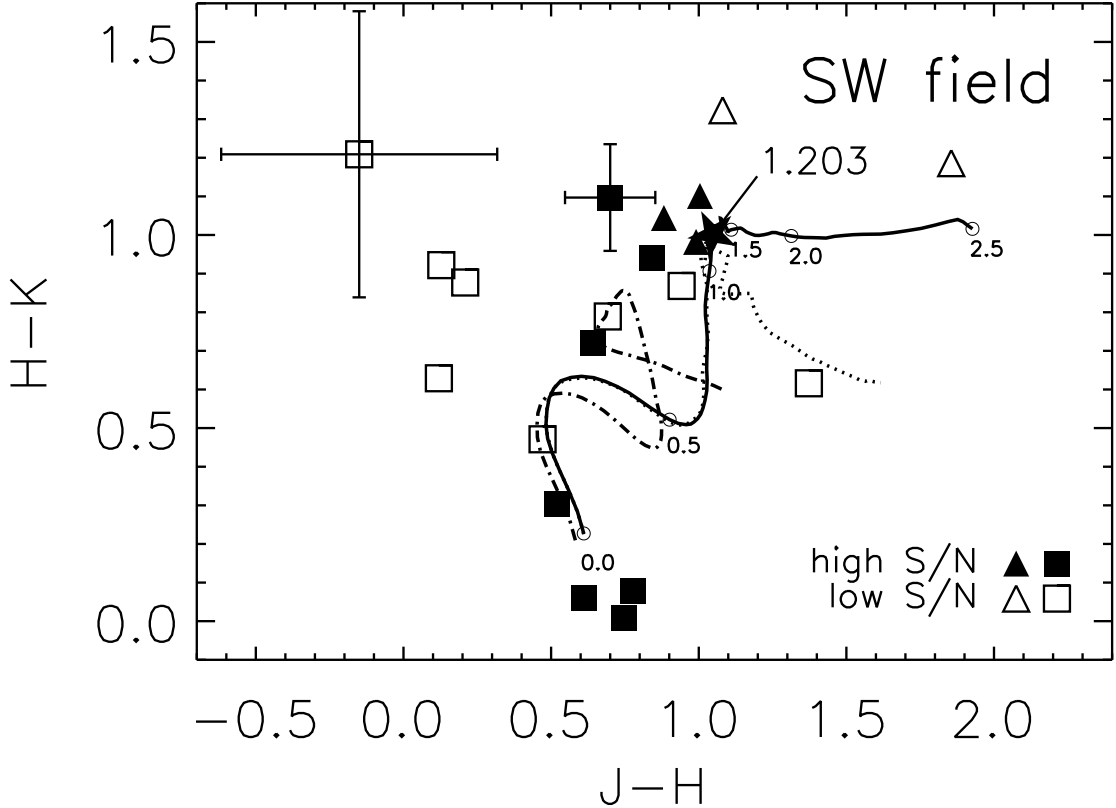}}
\caption[ero-phot.ps]{$JHK$ color-color plot for all objects in the
Keck/NIRC images SW of the quasar, the only field where we measured two
IR colors. The filled symbols are the data with errors in both colors of
$\le0.25$~mag, and the empty symbols are the lower S/N data. The
triangles indicate the $\Rs-K>5$ objects from
Figure~\ref{ero-colors}. The median errors for each subset are plotted
on one data point. The lines show the same theoretical models as in
Figure~\ref{ero-colors}. The observed colors of the well-detected red
galaxies suggest these objects lie at $z=1-2$.  The one object in the
field with a redshift, the emission-line ERO R6 at $z=1.203$, is marked
with a star ($\star$) and labeled.
\label{ero-colors-jhk-sw}}
\end{figure}

With the Keck and \HST\ findings, we now examine more closely the origin
of the colors of the 1213--0017 red galaxies and EROs.  Their colors are
consistent with $z>1$ ellipticals (Figures~\ref{cmd}
and~\ref{ero-colors}), and we have confirmed this inference for three of
them using Keck spectroscopy.  For these galaxies there is little doubt
the large $\Rs-K$ colors are due to old stellar populations with a
negligible amount of dust. For the red galaxies in the SW field, their
$\Rs JHK$ colors are also consistent with unreddened $z=1-2$ ellipticals
(Figures~\ref{ero-colors} and~\ref{ero-colors-jhk-sw}); the one SW
galaxy with a Keck spectrum, R6, supports this idea as it lies at
$z=1.203$, though its redshift is based on weak \oiin\ emission.  (The
spectrum has insufficent S/N to detect any continuum breaks, if
present.) So even if R6 is composed of an old, red stellar population
(its $\Rs$-band flux is only marginally brightened by the \oiin\ line),
the emission line points to a small amount of simultaneous
star-formation or weak AGN activity. In fact, R6 has quite a different
appearance in optical as compared to the near-IR
(Figure~\ref{zoom-contours}) which may suggest significant dust
reddening in its central regions, unlike most of the other red galaxies
(see below).

The other galaxy with a spectroscopic redshift, R7, has an optically
revealed AGN. Its $\Rs-K$ color is far too red compared to typical
Seyferts \citep[e.g.,][]{1994MNRAS.266..953K}. Its redness might come
from dust
or from old stars, although its optical spectrum shows no signs of the
latter. If the Balmer features in the spectrum are real, this would
suggest R7 has experienced significant star formation in the last few
$\sim10^8$ years and would hint that the red colors are due to dust
reddening.

For the remaining 1213--0017 red galaxies, we have no direct evidence
about their redshift nor their stellar content; all that we know is
their $\Rs-K$ colors.  One unlikely cause for their redness would be a
strong emission line redshifted into the $K$-band; the lowest possible
redshift for this to occur would be for H$\alpha$ at $z\approx2.1-2.6$.
It seems implausible that all or even any of these galaxies are at this
redshift given the very low surface density of strong high-redshift
H$\alpha$ emitters \citep[e.g.,][]{1996AJ....112.1794T,
1998ApJ...504..107B,1998ApJ...506..519T}.
In fact, the more likely case of emission-line contamination of the
broad-band magnitudes is in the $\Rs$-band, either from H$\alpha$ or
\oiin; more than half of $z=0.5-1.3$ field galaxies show \oiin\ emission
\citep{1997ApJ...481...49H}. In this case, the measured $\Rs-K$ colors
would be slightly bluer than the true color of the underlying stellar
continuum.

We find the observed optical and near-IR morphologies are fairly regular
and in good agreement with one another (Figure~\ref{zoom-contours}).
Both of these tell us that dust reddening in these galaxies is spatially
uniform on scales of $\lesssim4\ \hfperone$~kpc, as small as
$\approx0.9\ \hfperone$~kpc for the \HST\ images.  
This is in sharp contrast to the highly wavelength-dependent appearance
of the prototypical ERO HR~10: in the rest-frame near-UV/blue it has an
inverted S-shape while in the rest-frame red it shows a smooth compact
appearance \citep{1999ApJ...519..610D}. This implies HR~10 is an
interacting/disturbed system whose central region has a large amount of
dust, an interpretation augmented by its sub-mm emission which
presumably springs from a large amount of ongoing dusty star formation.
\HST\ optical morphologies for the first sub-mm selected sources, which
are possibly similar to HR~10, also show disturbed appearances
\citep{1998ApJ...507L..21S}.  Likewise, the rest-frame UV appearance of
local ultraluminous \IRAS\ galaxies (ULIRGs) are significantly different
than their appearance at longer wavelengths \citep{1999AJ....117.2152T}.

Therefore, we conclude the wavelength-independent appearance of the
1213--0017 red galaxies means they are unlikely to be dusty star-forming
galaxies like HR10 or ULIRGs. Though dust reddening may still still play
a role, most of their redness probably originates from the light of old
stars.  Dust variations could still exist on spatial scales below our
resolution limits.  For this reason, attempts to fit the broad-band SEDs
are potentially misleading given the uncertainties in the dust geometry
and distribution --- such fits would be highly degenerate.  Near-IR
images with comparable spatial resolution to the \HST\ optical data are
needed to address this possibility.

\subsubsection{Rest-Frame UV Spectral Features \label{uvspectra}}

Comparisons between the galaxy spectra and the \IUE\ stellar spectra
show the three continuum-break galaxies have real differences in their
rest-frame mid-UV spectra.  The spectra of R10 rises most rapidly to the
red and has the strongest 2640~\AA\ and 2900~\AA\ breaks, while R1's
continuum is noticeably flatter and its breaks are the weakest,
especially the 2900~\AA\ break.  R8's spectrum is intermediate between
R10 and R1.  Results from both cross correlation analysis
(\S~\ref{contbreak-spectra}) and examination by eye find R1 is
best-matched by early-type F~dwarfs, R8 by mid-type F~dwarfs, and R10 by
late-type F~and early-type G~dwarfs (Figure~\ref{absrestframe}) in this
wavelength region.  Furthermore, even when the \IUE\ spectra are
normalized only to the galaxy flux longward of 3000~\AA, there is still
good agreement between the star and galaxy spectra all the way to the
bluest wavelengths. This suggests any ongoing or recent star formation,
which would be revealed by a rising blue continuum, is small.

The UV continuum breaks at 2640~\AA\ and 2900~\AA\ are of special
interest. This region of an old galaxy's SED is expected to be dominated
by light from stars near the main-sequence turnoff.  As the stellar
population ages, the main-sequence turnoff mass decreases, and
later-type stars produce the UV SED, leading to larger break
amplitudes. In this fashion the rest-frame UV spectrum might be used to
age date the stellar population, or at least the portion contributing
the bulk of the UV flux. \citet{1997ApJ...484..581S} attempted to do
this in a robust fashion for the old galaxy LBDS 53W091 at
$z=1.55$. Given difficulties of absolute age calibration for these
features and their uncertain metallicity dependence
\citep{1998ApJ...492L.131H,yi99}, we do not attempt to age date the
populations of the 1213--0017 galaxies.

However, we can still study the stellar content of the spectroscopically
old 1213--0017 galaxies by comparing the strength of these breaks with
old populations in the local Universe. The spectra of R8 and R10 are
barely amenable to this line of investigation, but the S/N of R1's
spectrum is too low so we exclude it.  The standard indices for
measuring these breaks, 2609/2660 and 2828/2921 from
\citet{1990ApJ...364..272F}, are too narrow for our spectra: these use
$\approx$25~\AA\ wide bandpasses, meaning they include only 2--3
spectral resolution elements for our data. Instead, we use our own
broader custom indices, W2640 and W2900.  We also measure the
moderate-bandwidth $2600-3000$ index of \citet{1990ApJ...364..272F} to
track the UV continuum color.  These are all calculated in the standard
fashion, from the ratio of the fluxes in blue and red bandpasses and
expressed in magnitudes:
\begin{equation}
{\rm Index} \equiv -2.5 \times \log_{10}\left[
\frac{\bar{F_\lambda}({\rm blue})}
{\bar{F_\lambda}({\rm red})}\right],
\label{eqn-index}
\end{equation}
where $\bar{F_\lambda}$ is the average flux density in the
bandpasses. Objects with stronger breaks have larger values for the
indices.  The red bandpass of W2640 covers the entire top-hat feature
redward of the 2640~\AA\ break and ends before the Fe+Cr absorption at
2745~\AA; the blue bandpass has the same width and includes
\ion{Fe}{2}~$\lambda$2609 absorption and part of the broad
BL~$\lambda$2538 feature \citep[e.g., see][]{1998AJ....116.2297P}.  The
blue bandpass of W2900 spans the \mgii\ and \ion{Mg}{1}~$\lambda$2852
absorption, and the red bandpass is chosen to have the same width.
Definitions and measurements of the standard indices and our custom ones
are listed in Table~\ref{table-indices}.

\begin{figure}
\centering 
\includegraphics[angle=0,height=6.5in]{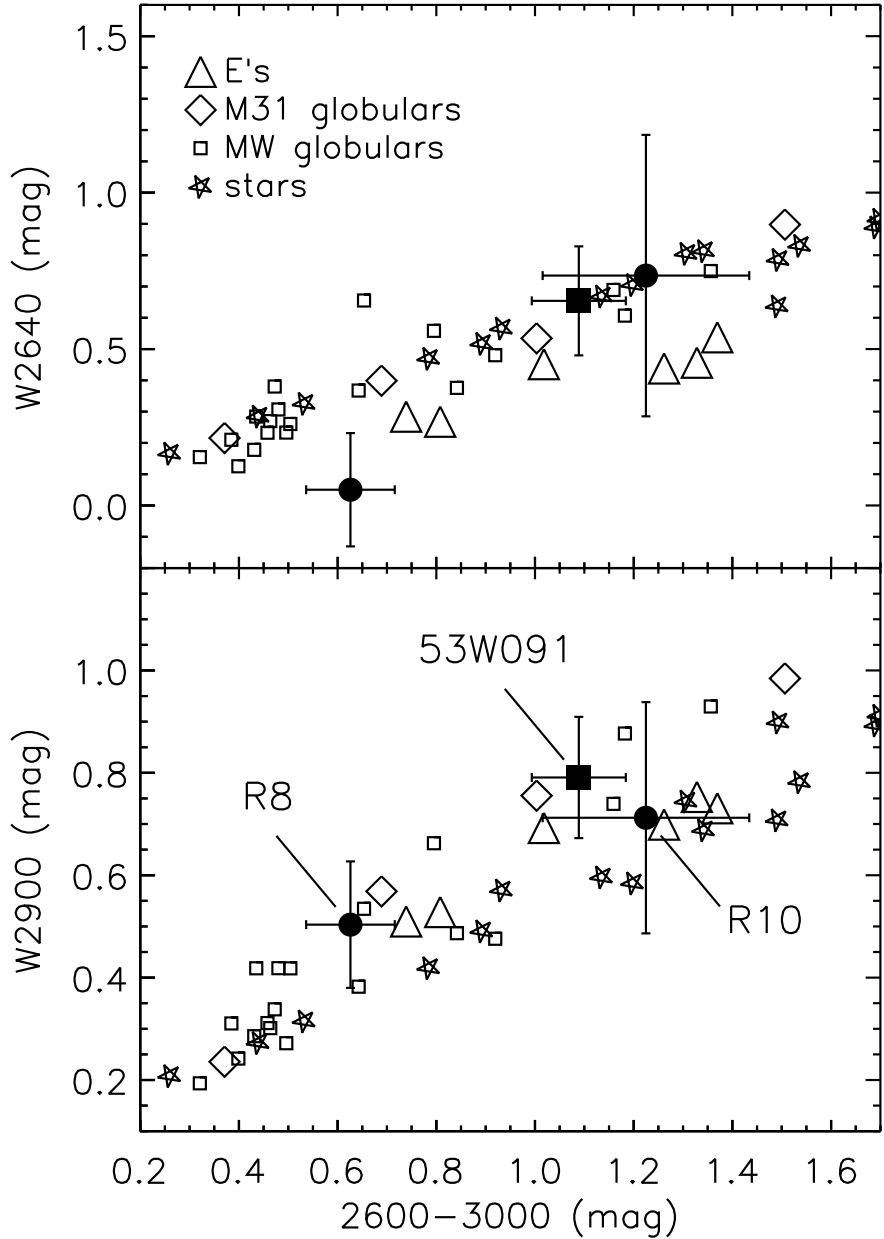}
\vskip 0.2in 
\caption{Strength of the UV-continuum breaks of red galaxies R8 and R10
compared with local old stellar populations and LBDS~53W091 at $z=1.55$.
W2640 and W2900 measure the 2640~\AA\ and 2900~\AA\ breaks,
respectively, and $2600-3000$ is a moderate-bandwidth UV color.  
Data for individual stars are from Fanelli \etal\ (1992), for Milky Way
globulars from Rose \& Deng (1998), and for M~31 globulars and
ellipticals from Ponder \etal\ (1998).  More metal-poor globulars lie to
the lower left and more metal-rich ones to the upper right.  R8 is more
similar to the more metal-poor local systems than R10.  The W2640 offset
between the M31 globular clusters and the local ellipticals is believed
to arise from the UV-excess population in the ellipticals.
\label{customindex}}
\end{figure}

Figure \ref{customindex} plots the results juxtaposed against individual
main-sequence stars \citep{1992ApJS...82..197F}, local old spheroids
\citep{1998AJ....116.2297P}, and the $z=1.552$ galaxy LBDS~53W091.  The
UV indices of R8 and R10 are consistent with local populations, though
the W2640 strengths of R8 and R10 differ, with the former having a
weaker 2640~\AA\ break than even the local elliptical sample.  Also, the
two galaxies have different $2600-3000$ colors, with R8 lying closer to
the metal-poor local populations, which have bluer mid-UV colors.  The
similar break amplitudes of LBDS~53W091 and R10 suggest these two
galaxies are comparably old.  If the analysis of 53W091's spectrum is
correct, its inferred age is barely compatible with present cosmological
parameters, even with the assumption of solar metallicity to reduce the
inferred age \citep{1997ApJ...484..581S}. Therefore, R10 must have had a
comparably high formation redshift, since the difference in look-back
time is only 5\% between $z=1.55$ and $z=1.31$, and also must have
experienced little or no recent star-formation.

For the local populations, the offset of the ellipticals from the stars
and globular clusters in plots of the 2640~\AA\ break versus $2600-3000$
color was noticed by \citet{1998AJ....116.2297P}, who suggested it was
due to the hot UV-excess (UVX) population in elliptical galaxies. The
UVX is thought to originate from post-main sequence stars, though its
specific origin remains uncertain \citep{oco99}. Interestingly, galaxy
R8 does appear to be more consistent with the local ellipticals than the
local non-UVX populations which might suggest a hot component in R8.  If
this could be confirmed for this galaxy and other high-redshift
ellipticals, a detailed study of their UV spectra might uncover the
origin of the UVX emission, since the potential UVX candidates evolve on
different time scales
\citep[e.g.,][]{1998ApJ...508L.139B,1999ApJ...513..128Y}.  However,
ongoing star formation, even just a small amount by mass, can greatly
brighten a galaxy's UV flux, thereby complicating such an analysis.



To summarize, comparison of the rest-frame UV spectra of R1, R8, and R10
shows a real scatter in their stellar content even though their observed
$K$-band fluxes, which should roughly trace galaxy mass, are within a
factor of two of each other. This may reflect differences in their star
formation and/or merging histories.  The UV color and break amplitudes
of R8 and R10 are consistent with $z=0$ old stellar populations, though
the data's limited S/N and spectral resolution hinders any more detailed
comparison, \eg, studying the weaker \mgiin\ and \ion{Mg}{1} absorption
features.  At a redshift of 1.3, we are observing these old galaxies
when the Universe was only 30\% of its present age. Therefore, improved
data, both in S/N and spectral resolution, could probe the evolving
composition of the oldest stellar populations from their relative youth
at high redshift to their advanced age at $z=0$.

\subsubsection{Relation to Galaxy Morphology and Implications for Formation}

The morphological results from the \HST\ images are somewhat surprising
if our suggestion that the 1213--0017 red galaxies contain old stars is
correct. Most of the galaxies do not seem to be early-type ones but
rather are disky or diffuse. Indeed, the one absorption-break galaxy
with \HST\ imaging, R10, looks like a compact edge-on disk --- this
galaxy does not resemble an elliptical galaxy morphologically even
though it does so spectroscopically.
These findings argue that not all old EROs are necessarily ellipticals,
at least in their structure. This may reflect a fundamental ambiguity in
identifying ellipticals at $z>1$ as compared to those at $z=0$:
are elliptical galaxies properly defined by their morphological or
spectrophotometric properties \citep[c.f.][]{1996MNRAS.283L.117K,sch99}?
If they formed by monolithic collapse at very high redshift,
there should be no ambiguity.  But in current scenarios of hierarchical
galaxy formation, massive early-type galaxies form late ($z\approx1-2$
depending on the cosmology) by agglomeration of smaller older sub-units,
so the stars of the resulting elliptical are actually much older than
the equilibrium morphology.  Simulations \citep{1998MNRAS.294..705K}
suggest these scenarios can still explain the homogeneity of the
early-type cluster galaxy population out to $z\approx0.9$
\citep[e.g.,][]{1993MNRAS.262..764A,1998A&A...334...99K,1998ApJ...501..571G,
1998ApJ...492..461S}. However, it is at $z\gtrsim1$ when differences are
predicted to become apparent, and there is little observational evidence
in this regime due to the few number of $z>1$ clusters known.
Spectroscopic follow-up of the remaining red galaxies in the 1213-0017
field combined with their \HST\ images should prove very useful in
addressing this issue.


\section{Conclusions}

We have found an overdensity of EROs in the field of the $z=2.69$ quasar
QSO~1213--0017 (UM~485), about a factor of 16 overdense compared to the
blank field ERO surface density and at least of factor of 6 overdense at
the 95\% confidence level.  The optical/IR colors of the EROs and
numerous other red galaxies in this field are consistent with those of
passively-evolving elliptical galaxies at $z>1$.  \HST\ optical imaging
shows a few of the red ($\Rs-K>5$) galaxies seem to have early-type
morphologies while the remainder are either disk galaxies or diffuse
objects without any obvious core.  Their near-IR morphologies are
consistent with those observed in the optical, unlike in the case of the
prototypical ERO HR~10. This suggests that the dust extinction in these
galaxies is either relatively smooth or that reddening by dust is not a
significant cause of the colors.




Follow-up Keck spectroscopy has measured redshifts for five red
galaxies.  Four lie at $z\approx1.31$, and three of these have
rest-frame UV absorption-line spectra similar to present-day elliptical
galaxies, making this the most distant concentration of old galaxies
spectroscopically confirmed to date. Including an \mgiin\ absorber seen
in the spectrum of the background quasar, there are five spectroscopic
redshifts at $z\approx1.31$ with a standard deviation of 1800~\kms\ and
a full range of 3800~\kms\ in the mean rest frame.

A number of lines of evidence possibly suggest the red galaxies in this
field delineate the presence of a massive high-redshift cluster.  The
ERO surface density is enhanced above blank field counts, and there are
five spectroscopic redshifts close together.  The reddest 1213--0017
galaxies, three of which have the spectra of early-type galaxies, are
nearly as luminous and as red as the brightest Coma cluster ellipticals.
The red galaxies also have a large angular extent on the sky, and a
considerable velocity spread, suggesting dynamical youthfulness.  An
impression of youthfulness is also conveyed by the roughly filamentary
distribution of the red galaxies on the sky.

Aside from the direct physical evidence, the presence of
spectroscopically old EROs provides circumstantial support to the idea
of a cluster.  Elliptical galaxies in the local Universe, presumably the
present-day counterparts of $z>1$ old EROs, are highly clustered and
predominantly found in high density regions
\citep[e.g.,][]{1980ApJ...236..351D}.  Such an analogy suggests that old
EROs should be found with others of their ilk --- this idea has yet to
be fully explored.
Since the hallmark of rich clusters is the presence of old ellipticals,
this $z=1.31$ field may be one of the most distant rich cluster
candidates to date. Moreover, regardless of whether this system is shown
to be a genuine massive cluster, the concentration of EROs is likely a
sign that this field is an uncommonly overdense region at high redshift,
which warrants follow-up studies.

Further observations are needed to develop a full physical picture of
this field.  A much larger sample of redshifts, inferred from multicolor
photometric redshifts and directly measured from deep spectroscopy, will
reveal the number of luminous galaxies at this redshift.  An expanded
spectroscopic sample will also be valuable to measure the galaxy
velocity distribution and to search for dynamical sub-structure.  X-ray
observations will allow us to look for hot intracluster gas, a sign of a
deep gravitational potential, and also for mass sub-structure. This
issue can also be addressed with radio observations to search for a
Sunyaev-Zeldovich decrement.
 
Extending galaxy cluster studies to $z>1$ has special importance since
current hierarchical formation theories predict massive galaxies
assemble from smaller sub-units during this epoch
\citep[e.g.,][]{1998MNRAS.294..705K}, so we can potentially test these
models by witnessing the formation process in situ.  Measuring the
high-redshift cluster abundance using wide-area surveys can provide
strong tests of cosmological models
\citep{1996MNRAS.282..263E,1997ApJ...485L..53B}.  However, even before a
large number of such clusters have been found, the few known to date are
valuable sites to investigate the processes which drive the evolution of
the oldest stellar populations, galaxies, and galaxy
clusters. Furthermore, these clusters can also be used as testing
grounds for $z>1$ cluster-search strategies, which by necessity are
derived from extrapolating the known properties of lower-redshift
clusters.  It may be that we have to abandon some common precepts formed
from studying local rich clusters in order to develop a complete
understanding of high-redshift clusters and their constituent galaxies
at an epoch of less than half the current age of the Universe.


\acknowledgments

Much of the data presented herein were obtained at the W.\ M.\ Keck
Observatory, which is operated as a scientific partnership among the
California Institute of Technology, the University of California and the
National Aeronautics and Space Administration.  The Observatory was made
possible by the generous financial support of the W.\ M.\ Keck
Foundation.  We are grateful to Hyron Spinrad, Aaron Barth, and Dave
Burstein for discussions which improved this work.  It is also a
pleasure for us to acknowledge Rychard Bouwens for the stellar
population model calculations, Adam Stanford for the redshifted Coma
color-magnitude relation, Dave Thompson for results from the CADIS
survey, and Jim Rose and Dave Burstein for providing UV spectra.  We
thank the NOAO staff for assistance with the Kitt Peak observing runs,
and Greg Wirth, David Sprayberry, and Ron Quick for supporting the
Keck/LRIS observations. Dan Stern and Andy Bunker provided helpful
advice and software for the LRIS reductions.  We also thank Emmanuel
Bertin for distributing SExtractor, and Wayne Landsman and Jonathan
Baker for contributing IDL routines.  Jenny Graves helped with the
initial near-IR reductions and photometry.  This research has made use
of the Digitized Sky Survey, NASA's Astrophysics Data System Abstract
Service, and the NASA/IPAC Extragalactic Database (NED).  A.~Dey
acknowledges financial support from NASA grant HF-01089.01-97A.  This
work has also received support from \HST\ NASA grant GO-06958.02-95A.

\clearpage






\clearpage
%

\thispagestyle{empty}

\begin{deluxetable}{lllcclcl} 
\rotate
\tablecaption{Spectroscopic Redshifts in the 1213--0017 Field \label{table-spectra}}
\tablewidth{0pt}
\tabletypesize{\small}

\tablehead{
\colhead{Identification} & \colhead{RA (J2000)} & \colhead{Dec (J2000)} & 
\colhead{$K$ (mag)} & \colhead{$\Rs-K$ (mag)\tablenotemark{a}} & \colhead{Morphology\tablenotemark{b}} & 
\colhead{$z$} & \colhead{Spectral Features}}

\startdata	                         
R1  & 12 15 43.2 & --00 35 39 & $18.30\pm0.07$ & $6.42\pm0.15$ & ---           & $1.317\pm0.005$ & B2640, \ion{Mg}{2}, B2900 \\
R6  & 12 15 48.1 & --00 34 37 & $18.89\pm0.07$ & $5.96\pm0.12$ & diffuse\tablenotemark{c} & $1.203\pm0.002$ & [O II] \\
R7  & 12 15 50.9 & --00 34 32 & $18.04\pm0.04$ & $5.84\pm0.07$ & compact+disk? & $1.319\pm0.002$ & [O II], C II], [Ne IV], C III]?\\
R8  & 12 15 44.8 & --00 34 25 & $18.94\pm0.09$ & $5.30\pm0.13$ & ---           & $1.298\pm0.004$ & B2640, \ion{Mg}{2}?, B2900, D4000? \\
R10 & 12 15 52.2 & --00 34 00 & $18.19\pm0.05$ & $6.26\pm0.08$ & edge-on disk  & $1.290\pm0.002$ & B2640, \ion{Mg}{2}, B2900, D4000?   \\
\\
Mg~II absorption\tablenotemark{d} & 12 15 49.8 & --00 34 34 &&&& 1.3196 &\\
                                  &            &            &&&& 1.5534 &\\
\enddata

\tablenotetext{a}{\Rs-band mags are on the AB system while $K$-band mags
are normalized to Vega (see \S~\ref{data-optical}). A flat spectrum
($f_\nu$ = constant) source has $\Rs-K=1.85$.}

\tablenotetext{b}{Morphologies from our {\sl HST} $F814W$ images. Sources with a blank line ``---'' were not observed.}

\tablenotetext{c}{Unlike the other EROs, which have similar morphologies
in the $F814W$ and $K$-band images, R6 appears diffuse in $F814W$ but
compact in the $K$-band images.  See \S~\ref{red-colors} and
Figure~\ref{zoom-contours}.}

\tablenotetext{d}{From Steidel \& Sargent (1992).}

\end{deluxetable}


\clearpage

\begin{deluxetable}{lcclcc}
\tablecaption{Rest-Frame UV Spectral Indices \label{table-indices}}
\tablewidth{0pt}
\tablehead{
\colhead{Name} & \colhead{Blue Passband (\AA)} & 
\colhead{Red Passband (\AA)} & \colhead{Definition} &
R8 & R10
}
\startdata
2600 -- 3000 & 2470--2670 & 2930--3130 & Fanelli \etal\ (1990) & $0.63\pm0.09$ & $1.22\pm0.21$ \\
2609/2660    & 2596--2623 & 2647--2673 &                       & $0.41\pm0.34$ & $0.62\pm0.61$ \\
2828/2921    & 2818--2838 & 2906--2936 &                       & $0.41\pm0.31$ & $0.29\pm0.36$ \\
&&&\\
W2640        & 2550--2625 & 2647--2722 & this work             & $0.05\pm0.18$ & $0.73\pm0.45$ \\
W2900        & 2773--2873 & 2910--3010 &                       & $0.50\pm0.12$ & $0.71\pm0.23$ \\
\enddata




\end{deluxetable}


\end{document}